\documentclass[aip,jcp,reprint,twocolumn,amsmath,amssymb,amsfonts,floatfix,letterpaper]{revtex4-1}
\usepackage{mathptmx}
\usepackage[scaled=0.90]{helvet}
\usepackage{courier}
\normalfont

\usepackage{miniacs}
\setkeys{acs}{usetitle = true}

\usepackage{graphicx}
\usepackage{braket}
\usepackage{color}
\usepackage{simplewick}
\usepackage{mathptmx}
\usepackage{booktabs}

\newcommand{\peq}{\;+\!\!=}
\newcommand{\sosep}{,\,} 
\newcommand{\mat}[1]{\mathbf{#1}}
\newcommand{\groupname}[1]{#1} 
\newcommand{\groupelement}[1]{\hat{#1}} 
\newcommand{\ti}[2]{\;{#1}{{[#2]}}\;} 

\newcommand{\sign}{\operatorname{sign}}
\newcommand{\canfn}{z}

\renewcommand{\footnote}[1]{{\color{green}[#1]}}

\begin{document}

\title{{\color{blue}A simple permutation group approach to spin-free higher-order coupled-cluster methods}\vspace*{0.18cm}}
\author{Cong Wang}
\affiliation{The Pennsylvania State University; 401A Chemistry Building; University Park, PA 16802 (USA)}
\author{Gerald Knizia}
\email{knizia@psu.edu}
\affiliation{The Pennsylvania State University; 401A Chemistry Building; University Park, PA 16802 (USA)}

\date{\today}

\begin{abstract}
  We present a general-order spin-free formulation of the single-reference closed-shell coupled-cluster method.
 We show that the working equations of a fully biorthogonal contravariant projection formulation of the residual equations, as near-universally used in closed-shell CCSD, can also be defined at the CCSDT and CCSDTQ levels, despite singularities in the spin projection manifolds.
 We describe permutation-group based techniques for obtaining and simplifying the equations
 encountered in general second-quantization-based methods; this includes a permutation group based approach of evaluating second-quantized matrix elements into tensor contraction networks, and the use of Portugal's \emph{double coset canonical representation} technique [Int. J. Mod. Phys. C 13, 859 (2002)] for eliminating redundant terms.
A computer implementation of our techniques is simple, because no operator-valued symbolic algebra is required.
Explicit working equation lists for closed-shell CCSD, CCSDT, and CCSDTQ in the semi-biorthogonal formulation are provided.
We also release open-source computer programs for both deriving and numerically evaluating these equations.
\end{abstract}
\maketitle

\section{Introduction}

Coupled-cluster (CC) theory \cite{Cizek:66,cc:07,Bartlett:07,bartlett2009many,bartlett2012coupled} is one of the most successful frameworks for accurately describing many-electron
correlations. To date, the CC method has spread to almost every aspect of modern electronic structure theory, including open-shell systems \cite{datta2014analytic},
treatment of excited states \cite{kinoshita2005coupled,hino2006tailored,shiozaki2008equations,musial2011charge,datta2012multireference,jagau2012linear,samanta2014excited,eriksen2014equation},
multi-reference approaches \cite{evangelista2006high,prochnow2009analytic,evangelista2010perturbative,musial2011multireference,musial2011multi,demel2013additional,lyakh2011multireference,
evangelista2011orbital,evangelista2012sequential,datta2011state,datta2011spin,das2012superior,maitra2012unitary,koehn2013state,nooijen2014communication,rolik2014quasiparticle,maitra2014aspects,melnichuk2014relaxed},
strong correlation \cite{piecuch2005renormalized,kats2013communication,small2014coupled,bulik2015can,henderson2015pair,rishi2016assessing},
reduction of computational scaling \cite{huntington2012accurate,rolik2013efficient,parrish2014communication,liakos2015domain,werner2015scalable},
accelerating basis set convergence via F12 terms \cite{shiozaki2008explicitly,shiozaki2009higher,shiozaki2009explicitly,knizia2009simplified}.
Additionally, combinations with other computational methods\cite{eriksen2014lagrangian,eriksen2015convergence}, CC-based energy decomposition analysis\cite{azar2012energy}, massive parallel implementations\cite{solomonik2014massively} and even solid-state extensions versions\cite{mcclain2016spectral,mcclain2017gaussian} have been reported.
Progress is also being made regarding the formal structure of CC; for example,
various forms of the exponential ansatz\cite{nooijen2000can,nakatsuji2000structure,van2001two,kutzelnigg2005minimal}
and the approximations of the Baker-Campbell-Hausdorff (BCH) expansion \cite{evangelista2012approximation} have been investigated.

However, it has also become clear that even for electronically rather benign molecular systems,
the ``gold standard'' CCSD(T) at its basis set limit cannot be expected to reliably reach sub-chemical accuracy
($\leq$ 1 kcal mol$^{-1}$) in relative energies.\cite{karton2006w4,karton2010performance,karton2011w4,Stanton:04,harding2011towards,karton2016computational}
This level of accuracy can be essential for predicting reaction mechanisms\cite{plata2015case} and crystal polymorphs.\cite{yang2014ab}
To this end, high-order coupled-cluster methods such as CCSDT and CCSDTQ must be invoked.\cite{karton2006w4,Stanton:04}
These are the subject of the current article.

While the equations for general-order coupled-cluster method have been obtained by several authors,\cite{Harris:99,Olsen:00,Hirata:00,Kallay:00,Kallay:01,engels2011fully} most of the previous derivations are based on a spin-orbital formulation.
It is well-known that for closed-shell electronic systems, a formulation based on (spin-free) spatial orbitals, rather than spin orbitals, and combined with a biorthogonal projection to define the CC matrix elements, could potentially reduce the computational cost by a large prefactor.\cite{hampel1992comparison}
This results from two factors: First, in a spatial orbital formulation, there are less wave function amplitudes than in a spin-orbital formulation. And second, the tensors representing these amplitudes have a significantly simpler permutational symmetry structure in the spatial orbital case; this structure is more amendable to highly efficient matrix-multiplication based computational kernels than the anti-symmetric tensors used in spin-orbital methods.
These aspects were recently exploited by Matthews, Gauss, and Stanton,\cite{Stanton:14,Stanton:15,springer2017spin,eriksen2015convergence,eriksen2016assessment,matthews2016high}
who developed highly efficient CC methods up to CCSDTQ, based on a non-orthogonal spatial-orbital formulation.
Nevertheless, on the formal side, their scheme still employs spin orbitals and standard diagrammatic expansions in the definition of the CC matrix elements, and introduces the spatial orbitals by combining them.

In the present article, we shall introduce an alternative single-reference orbital-based coupled-cluster formulation for closed-shell systems. We aim at reducing the prefactor of the higher-order CC without any approximation. Our objectives are fourfold:
\begin{enumerate}
   \item[(i)] employing spin-free excitation operators and spatial orbitals directly to derive the working equations, rather than recasting a spin-orbital formulation into spatial orbitals;
   \item[(ii)] extending the biorthogonal contravariant projection beyond the double excitation,\cite{pulay1984efficient,koch1990coupled,hampel1992comparison,knowles1993coupled} which could accelerate the convergence of perturbative updating and provide a future route into perturbative corrections to iterative high-order CC methods;
   \item[(iii)] fully eliminating the redundancy of the spin-free parametrization beyond the double excitation, which greatly reduces the number of terms in the working equations;
   \item[(iv)] at the symbolic computation level, adopting unified permutation group and double-coset representation \cite{butler:GroupHomomorphisms,portugal:DoubleCosetRep1,martingarcia:InvarPackage,martingarcia:xperm} techniques from computational group theory to construct a method for deriving and symmetrizing the coupled cluster working equations,
   which is fully algebraic and well suited for a computer implementation.
\end{enumerate}

The targets (i) - (iii) are rather specific to the CC method. The point (iv) is applicable to general second-quantization-based methods.

Before proceeding, we shall first recapitulate the formal framework of {\it Quantum Chemistry in Fock Space}, \cite{kutzelnigg:QCinFockSpace1,kutzelnigg:QCinFockSpace2,kutzelnigg:QCinFockSpace3,kutzelnigg1985quantum,Kutzelnigg1989,kutzelnigg:GeneralizedNormalOrder} on which the current work is based.
In particular, we reiterate the concepts of generalized normal ordering\cite{kutzelnigg:GeneralizedNormalOrder} and spin-free excitations.
Based on these techniques, we then discuss our approach to the spin-free version of closed-shell CC in pursuit of objective (i).
To address (ii) and (iii), we then discuss a semi-biorthogonal formulation of the CC residual equations for excitation levels beyond singles \& doubles, and permutation group based methods of eliminating redundant equations.
Next, the direct evaluation of coupling coefficients (DECC) method is described as a simple and straightforward way of evaluating the second-quantization matrix elements into tensor contraction networks---depending on perspective, this technique can be seen as a variant, extension, or alternative to the standard diagrammatic and generalized Wick-theorem based methods of performing this task.
Finally we briefly explain our algorithms and describe the open-source Python and C++ programs implementing the proposed techniques.

\section{Tensor notation and normal ordering}
We employ the formalism of {\it Quantum Chemistry in Fock Space}.
\cite{kutzelnigg:QCinFockSpace1,kutzelnigg:QCinFockSpace2,kutzelnigg:QCinFockSpace3,kutzelnigg1985quantum,Kutzelnigg1989,kutzelnigg:GeneralizedNormalOrder}
The tensor notation is used. We also adopt the Einstein summation convention: namely, repeated indices in a tensor expression imply summation, except for the indices $\{ \sigma_1, \cdots, \sigma_k \}$ which represent spin components.
General spatial orbitals are indexed with $R$, $S$, $T$,\ldots, occupied spatial orbitals with $I$, $J$, $K$,\ldots, and virtual orbitals with $A$, $B$, $C$,\ldots.
An $\hat E$-operator is defined as the spin-summed substitution
\begin{align}
\hat E_{R_1 \cdots R_k }^{S_1 \cdots S_k} := \sum_{\sigma_1, \cdots, \sigma_k\in\{\alpha,\beta\}} \hat e_{R_1 \sigma_1\sosep\cdots\sosep R_k \sigma_k }^{S_1 \sigma_1\sosep\cdots\sosep S_k \sigma_k}, \label{eq:SpinFreeEop}
\end{align}
where $\hat e_{R_1 \sigma_1\sosep\cdots\sosep R_k \sigma_k }^{S_1  \sigma_1\sosep\cdots\sosep S_k \sigma_k}$ are the normal-ordered spin-orbital substitution operators.\cite{kutzelnigg:GeneralizedNormalOrder}
The normal ordering is defined with respect to the reference state $\ket{\Phi}$,
which in our present scope is a closed-shell Slater determinant (such as obtained by Hartree-Fock or Kohn-Sham).
With this, the explicit expressions for the first two normal ordered spin-orbital substitution operators\cite{kutzelnigg:GeneralizedNormalOrder} (called $\hat e$-operators in the following) are
\begin{align}
\hat e_{R_1 \sigma_1}^{S_1 \sigma_1} &:= \hat a_{R_1 \sigma_1 }^{S_1 \sigma_1} - \gamma_{R_1 \sigma_1 }^{S_1 \sigma_1}, \label{e1}
\\
\hat e_{R_1 \sigma_1\sosep R_2 \sigma_2 }^{S_1 \sigma_1\sosep S_2 \sigma_2 } &:=
\hat a_{R_1 \sigma_1\sosep R_2 \sigma_2 }^{S_1 \sigma_1\sosep S_2 \sigma_2 }
- \Big(\gamma_{R_1 \sigma_1 }^{S_1 \sigma_1} \hat a^{S_2 \sigma_2 }_{R_2 \sigma_2}
+ \gamma^{S_2 \sigma_2  }_{R_2 \sigma_2 } \hat a^{S_1 \sigma_1 } _{R_1 \sigma_1}
\notag\\&\qquad- \gamma_{R_2 \sigma_2}^{S_1 \sigma_1} \hat a^{S_2 \sigma_2}_{R_1 \sigma_1}
- \gamma^{S_2 \sigma_2}_{R_1 \sigma_1 } \hat a^{S_1  \sigma_1 }_{R_2 \sigma_2}
- \gamma^{S_1 \sigma_1\sosep S_2 \sigma_2}_{R_1 \sigma_1\sosep R_2 \sigma_2}\Big). \label{e2}
\end{align}
The employed $k$-electron reduced density matrix (RDM) of the reference state is defined as
\begin{align}
\gamma_{R_1 \sigma_1\sosep R_2 \sigma_2 \sosep\cdots\sosep R_k \sigma_k }^{S_1\sigma_1\sosep S_2\sigma_2\sosep \cdots\sosep S_k \sigma_k }:=\Braket{\Phi|\hat{a}_{R_1 \sigma_1\sosep R_2 \sigma_2\sosep \cdots\sosep  R_k \sigma_k }^{S_1\sigma_1\sosep S_2\sigma_2\sosep \cdots\sosep S_k \sigma_k}|\Phi}\label{eq:SpinOrbitalRdm}
\end{align}
and the elementary spin-orbital substitution operator
\begin{align}
\hat a_{R_1 \sigma_1\sosep R_2 \sigma_2\sosep \cdots\sosep  R_k \sigma_k }^{S_1\sigma_1\sosep S_2\sigma_2\sosep \cdots\sosep S_k \sigma_k} := \hat{a}_{S_1 \sigma_1}^{\dagger} \cdots \hat a_{S_k \sigma_k}^{\dagger} \hat a_{R_k\sigma_k} \cdots \hat a_{R_1\sigma_1}\label{eq:SpinOrbSubstOp}
\end{align}
denotes a string of elementary creation and destruction operators with respect to the genuine vacuum.

The (recursive) definition for higher-order normal-ordered $\hat e$-operators beyond Eqs.~\eqref{e1} and \eqref{e2} is given in Ref.~\onlinecite{kutzelnigg:GeneralizedNormalOrder}, but it is not needed in the present article:
Note that the reference-RDMs defined by Eq.~\eqref{eq:SpinOrbitalRdm} vanish whenever any of the involved indices $\{R_k\}$ or $\{S_k\}$ refers to a virtual orbital.
As a consequence, in the special cases of pure excitation or de-excitation operators (i.e., operators which \emph{exclusively} excite occupied orbitals of the reference function into virtual orbitals, or vice versa), the normal ordered ($\hat e$) and elementary ($\hat a$) substitution operators are identical:
\begin{align}
 \hat e_{I_1 \sigma_1}^{A_1 \sigma_1} &= \hat a_{I_1 \sigma_1}^{A_1 \sigma_1} - \gamma_{I_1 \sigma_1}^{A_1 \sigma_1} =  \hat a_{I_1 \sigma_1}^{A_1 \sigma_1}, \label{eai1} \\
\hat e_{I_1 \sigma_1\sosep I_2 \sigma_2 }^{A_1 \sigma_1\sosep A_2 \sigma_2 } &=
\hat a_{I_1 \sigma_1\sosep I_2 \sigma_2 }^{A_1 \sigma_1\sosep A_2 \sigma_2 }
- \Big(\gamma_{I_1 \sigma_1 }^{A_1 \sigma_1} \hat a^{I_2 \sigma_2 }_{A_2 \sigma_2}
+ \gamma^{A_2 \sigma_2  }_{I_2 \sigma_2 } \hat a^{A_1 \sigma_1 } _{I_1 \sigma_1}
\\&\qquad- \gamma_{I_2 \sigma_2}^{A_1 \sigma_1} \hat a^{A_2 \sigma_2}_{I_1 \sigma_1}
- \gamma^{A_2 \sigma_2}_{I_1 \sigma_1 } \hat a^{A_1  \sigma_1 }_{I_2 \sigma_2}
- \gamma^{A_1 \sigma_1\sosep A_2 \sigma_2}_{I_1 \sigma_1\sosep I_2 \sigma_2}\Big) \nonumber \\
&= \hat a_{I_1 \sigma_1\sosep I_2 \sigma_2 }^{A_1 \sigma_1\sosep A_2 \sigma_2 }, \quad(\ldots)\label{eai2}
\end{align}
For this reason, the normal ordering does not affect the actual cluster-operators or residual projections (\emph{vide infra}) in the single-reference CC methods treated here.

However, advantages of normal ordering are obtained in the representation of the Hamiltonian, which in terms of $\hat E$ operators takes the form:\cite{kutzelnigg2010spinfree}
\begin{align}
\hat H &= E_{\mathrm{ref}} + f_R^S \hat E_S^R + \frac{1}{2} W_{RS}^{TU} \hat E^{RS}_{TU}, \label{Hop}
\\ E_{\mathrm{ref}} &= \braket{\Phi|\hat H|\Phi} = 2 h^I_I + 2 W_{JI}^{JI} - W^{IJ}_{JI},
\\ f^R_S &= h^R_S + 2 W_{RI}^{SI} - W^{RI}_{IS}. \label{Fock}
\end{align}
Thus, (i) the reference energy $E_{\mathrm{ref}}$ is separated and the Hartree-Fock mean field $\hat f$ explicitly appears as a one-particle operator, and (ii) internal contractions within the indices of the Hamiltonian are avoided.
Point (ii) is due to the fact that the effects from the internal contractions have been cancelled by the definitions of normal ordering in Eqs.~\eqref{e1} and \eqref{e2}.

\section{Spin-Free Coupled Cluster Theory}
\subsection{Spin-free CC Equations}
Coupled-cluster theory parameterizes a correlated $N$-electron state $|\Psi\rangle$ via an exponential wave operator $\exp(\hat T)$ applied to a reference state $\ket{\Phi}$. In our present scope, $| \Phi \rangle$ is a closed-shell Slater determinant.
For $K$th order CC, this wave function ansatz reads
\begin{align}
| \Psi \rangle &= \exp(\hat T) | \Phi \rangle, \\
\hat T &= \hat T_1 + \hat T_2 + \cdots + \hat T_K, \label{eq:Top} \\
\hat T_k &=  \frac{1}{k!}     t_{  A_1 \cdots A_k }^{I_1 \cdots I_k } \hat E_{I_1 \cdots I_k }^{A_1 \cdots A_k },  \qquad(k=1,2,3,\cdots K) \label{t}
\end{align}
and in this expression, the cluster amplitudes $t_{  A_1 \cdots A_k }^{I_1 \cdots I_k }$ are the quantities to be determined.
If the cluster operators $\hat T_1,\ldots,\hat T_K$ up to $K=N$ (number of electrons) are included, this ansatz is capable of recovering the exact $N$-electron ground state wave function of $\hat H$ (as well as any other $N$-electron wave function).
In practice, the expansion in Eq.~\eqref{eq:Top} is truncated to lower order, with typical $N$ being 2 (CCSD), 3 (CCSDT), or 4 (CCSDTQ).

In the standard coupled-cluster theory, both the CC energy and the cluster amplitudes $t_{A_1 \cdots A_k }^{I_1 \cdots I_k}$ are determined via projections of the time-independent Schr\"{o}dinger equation
\begin{align}
    \hat H | \Psi \rangle &= E | \Psi \rangle
\\ \Rightarrow \hat H \exp(\hat T) | \Phi \rangle &= E \exp(\hat T) | \Phi \rangle.\label{eq:CcSchroedinger}
\end{align}
For a $\hat T$ of limited order, Eq.~\eqref{eq:CcSchroedinger} can generally not be fulfilled exactly.
However, it can be fulfilled in a limited subspace of the full Fock space, and thereby used to construct conditions defining a unique set of cluster amplitudes $t^{IJ\cdots}_{AB\cdots}$.
Concretely, left-multiplying Eq.~\eqref{eq:CcSchroedinger} by $\exp(-\hat T)$ and projecting onto the space spanned by $\bra{\Phi}$ and $\bra{\Phi} \hat E^{IJ\cdots}_{AB\cdots}$ yields
\begin{align}
 E &= \braket{\Phi | \exp(-\hat T) \hat H \exp(\hat T)|\Phi}, \label{eq:CcEnergySimTrans}
\end{align}
as an energy equation, and a set of residual equations
\begin{align}
   r^{AB\cdots}_{IJ\cdots} &= \braket{\Phi|\hat E^{IJ\cdots}_{AB\cdots}  \exp(-\hat T) \hat H \exp(\hat T)| \Phi} = 0, \label{eq:CcResidualEqsimTrans}
\end{align}
which (implicitly) determine the cluster amplitudes $t^{IJ\cdots}_{AB\cdots}$ as the set of unknowns for which $r^{AB\cdots}_{IJ\cdots}=0$.
It can be shown that Eqs.~\eqref{eq:CcEnergySimTrans} and \eqref{eq:CcResidualEqsimTrans} are equivalent to
\begin{align}
 E &= \braket{\Phi | \hat{H} \exp(\hat{T})|\Phi}_c,  \label{eq:CcEnergy1} \\ 
r^{AB\cdots}_{IJ\cdots} &= \braket{\Phi|\hat{E}^{IJ\cdots}_{AB\cdots} \hat{H} \exp(\hat{T})| \Phi}_c = 0
 \label{eq:CcResidualEqFormal},   
\end{align}
where the subscript $c$ denotes that only connected terms are retained in the matrix elements.\cite{bartlett2009many}

\subsection{Biorthogonal contravariant projections}\label{sec:BiorthProj}
In Eq.~\eqref{eq:CcResidualEqFormal} it is not essential that the residuals are defined by projecting the Schr\"{o}dinger equation Eq.~\eqref{eq:CcSchroedinger} onto $\bra{\Phi} \hat E^{IJ\cdots}_{AB\cdots}$ directly.
A new set of residual equations $\tilde{r}^{AB\cdots}_{IJ\cdots}=0$ will define identical cluster amplitudes if the $\tilde{r}^{AB\cdots}_{IJ\cdots}$ are obtained by any non-singular transformation of the set of $\{r^{AB\cdots}_{IJ\cdots}\}$.
In practice, it has been found that rather than using $\bra{\Phi} \hat E^{IJ\cdots}_{AB\cdots}$ directly, certain linear combination of de-excitation operators can substantially reduce the complexity of the resulting residual equations.
For instance, the modified de-excitation
\begin{align}
\tilde{E}^{IJ}_{AB} &= \frac{1}{6} \left(   2 \hat E^{IJ}_{AB}  +   \hat E^{JI}_{AB}    \right)  \label{tilde}
\end{align}
reduces the number of working equations and establishes a one-to-one correspondence between the residual $r_{AB}^{IJ}$ and amplitude $t_{AB}^{IJ}$.  It has been used in efficient implementations of the CCSD method. \cite{pulay1984efficient,koch1990coupled,hampel1992comparison,knowles1993coupled}
However, previous attempts\cite{schutz2013orbital,Stanton:15} to generalizing the contravariant de-excitations to higher than double-excitations have found this task to be not-trivial.\cite{Helgaker:00r}
In Sec.~\ref{sec:BiorthHigherORders} we discuss these difficulties and introduce a scheme to work around this problem.

\subsection{Semi-biorthogonal contravariant projections for triples, quadruples, and higher}
\label{sec:BiorthHigherORders}

Let us first discuss the source of Eq.~\eqref{tilde}.
The definition in Eq.~\eqref{tilde} is obtained by demanding that $\bra{\Phi}\tilde{E}^{IJ}_{AB}$ and $\bra{\Phi}\tilde{E}^{JI}_{AB}$ are a bi-orthogonal basis for $\hat E^{AB}_{IJ}\ket{\Phi}$ and $\hat E^{AB}_{JI}\ket{\Phi}$, i.e., that
\begin{align}
   \braket{\Phi|\tilde{E}^{IJ}_{AB} \hat E^{AB}_{IJ}|\Phi} &= 1
 & \braket{\Phi|\tilde{E}^{JI}_{AB} \hat E^{AB}_{IJ}|\Phi} &= 0 \notag
\\ \braket{\Phi|\tilde{E}^{IJ}_{AB} \hat E^{AB}_{JI}|\Phi} &= 0
 & \braket{\Phi|\tilde{E}^{JI}_{AB} \hat E^{AB}_{JI}|\Phi} &= 1.
   \label{eq:DoublesBiorth}
\end{align}
While this demand technically only guarantees simple expressions for overlap matrices, we note that
also residual equations like $\tilde{r}^{AB\cdots}_{IJ\cdots} = \braket{\Phi|\tilde{E}^{IJ\cdots}_{AB\cdots}  \exp(-\hat T) \hat H \exp(\hat T)| \Phi}$
can be regarded as overlap matrix elements between the vectors ``$\bra{\Phi}\tilde{E}^{IJ}_{AB}$" and ``$\exp(-\hat T) \hat H \exp(\hat T)\ket{\Phi}$",
and therefore simpler residual equations can be expected as well.

To realize Eq.~\eqref{eq:DoublesBiorth}, $\tilde{E}^{IJ}_{AB}$ must be a linear combination of excitation operators differing only in label permutations, i.e.,
\begin{align}
   \tilde{E}^{IJ}_{AB} = v_1 \hat E^{IJ}_{AB} + v_2 \hat E^{JI}_{AB}\label{eq:DoublesBiOrthCoeffs},
\end{align}
since only $\bra{\Phi}\hat E^{IJ}_{AB}$ and $\bra{\Phi}\hat E^{JI}_{AB}$ can have non-zero overlap with $\hat E^{AB}_{IJ}\ket{\Phi}$ and $\hat E^{AB}_{JI}\ket{\Phi}$.
Taking this into account, the actual coefficients of Eq.~\eqref{tilde} can be computed via linear algebra in the space of label permutations, as suggested by Schaefer and coworkers.\cite{koch1990coupled}
Namely, if we define the ``permutational overlap matrix''
\begin{align}
\mat M:= \begin{bmatrix}
  \langle \Phi |   \hat E_{AB}^{IJ} \hat E_{IJ}^{AB} | \Phi \rangle &  \langle \Phi |   \hat E_{AB}^{JI} \hat E_{IJ}^{AB} | \Phi \rangle \\
   \langle \Phi |   \hat E_{AB}^{IJ} \hat E_{JI}^{AB} | \Phi \rangle &    \langle \Phi |   \hat E_{AB}^{JI} \hat E_{JI}^{AB} | \Phi \rangle
  \end{bmatrix}  =  \begin{bmatrix}
  4 &  -2 \\
 -2 &  4
  \end{bmatrix},\label{eq:MmatrixDoubles}
\end{align}
then Eqs.~\eqref{eq:DoublesBiorth} and \eqref{eq:DoublesBiOrthCoeffs} can be cast into the form
\begin{align}
   \begin{bmatrix}
   v_1 & v_2
\\ v_2 & v_1
   \end{bmatrix}  \mat M \begin{bmatrix} 1 & 0 \\ 0 & 1 \end{bmatrix}= \begin{bmatrix} 1 & 0 \\ 0 & 1 \end{bmatrix},\label{eq:DoublesContravariantProjectionAlgebraic0}
\end{align}
or, equivalently (note $\mat M = \mat M^T$),
\begin{align}
\mat M \begin{bmatrix}
   v_1
\\ v_2
   \end{bmatrix} = \begin{bmatrix} 1 \\ 0 \end{bmatrix}.\label{eq:DoublesContravariantProjectionAlgebraic}
\end{align}
Eq.~\eqref{eq:DoublesContravariantProjectionAlgebraic} is solved by $\begin{bmatrix} v_1 & v_2\end{bmatrix}=\frac{1}{6} \begin{bmatrix} 2 & 1\end{bmatrix}$, recovering Eq.~\eqref{tilde}.

Along this line, we also obtain
\begin{align}
\tilde{E}^{I}_{A} = \frac{1}{2} \hat E^{I}_{A}   \label{tilde0}
\end{align}
for the ``bi-orthogonal'' single excitation operators, which differ from the regular operators only in normalization.

Unfortunately, this construction cannot be directly be extended to higher orders than double substitutions.
While permutational overlap matrices $\mat M$ for triples (with order $3!=6$), quadruples (with order $4!=24$), etc., can still be straightforwardly defined in analogy with Eq.~\eqref{eq:MmatrixDoubles}, these $\mat M$ matrices are singular.
Therefore no true bi-orthogonal projections fulfilling the generalization of the equation system \eqref{eq:MmatrixDoubles}, and its algebraic reformulation 
\begin{align}
\mat M 
\begin{bmatrix}
   v_1
\\ v_2
\\ \vdots
\\ v_{N!}
\end{bmatrix}  = \begin{bmatrix} 1 \\ 0 \\ \vdots \\ 0 \end{bmatrix},\label{eq:ContravariantProjectionAlgebraicN}
\end{align}
exist in the case of substitution degrees $N$ larger than two. [It is noteworthy that this only applies to spin $1/2$-particles like electrons; 
however, we observed that if hypothetical fermions of spin $2/2$ would exist, their triples $\mat M$ matrix would be non-singular, allowing for true bi-orthogonal triple-de-excitations, too;
similarly, for hypothetical fermions of spin $3/2$, also true bi-orthogonal quadruple de-excitations can be defined, and so on.]
Nevertheless, even in these cases of substitution degree $N\geq 3$ we can still look for the projective linear combinations $\mat v = \begin{bmatrix} v_1 & v_2 & \ldots & v_{N!}\end{bmatrix}^T$ which fulfill the bi-orthogonality conditions 
Eq.~\eqref{eq:ContravariantProjectionAlgebraicN}
as closely as possible in a linear least squares sense.
Hence, we adopt the linear least squares solution of
\begin{align}
   \left\Vert\left(
\mat M 
\begin{bmatrix}
   v_1
\\ v_2
\\ \vdots
\\ v_{N!}
\end{bmatrix} - \begin{bmatrix} 1 \\ 0 \\ \vdots \\ 0 \end{bmatrix}
   \right)\right\Vert_2\rightarrow \min \label{pseudo}
\end{align}
to define the ``semi-biorthogonal'' linear combinations for the higher orders.
Additionally, we scale the resulting linear combinations of permutations such that $(\alpha \mat v^T) \mat M \begin{bmatrix} 1 & 0 & 0 & \ldots\end{bmatrix}^T=1$; this achieves that
\begin{align}
   \braket{\Phi_0|{\tilde E}^{IJK\ldots}_{ABC\ldots} E^{ABC\ldots}_{IJK\ldots}|\Phi_0}=1
\end{align}
after the scaling.
Executing this process, we obtain the following ``semi-biorthogonal'' de-excitation operator
\begin{align}
\tilde{E}^{IJK}_{ABC} =& \frac{17}{120} \hat E^{IJK}_{ABC}  - \frac{1}{120} \hat E^{IKJ}_{ABC} - \frac{1}{120}   \hat E^{JIK}_{ABC}
\notag\\&- \frac{7}{120}   \hat E^{JKI}_{ABC}      -\frac{7}{120}   \hat E^{KIJ}_{ABC}   -\frac{1}{120}   \hat E^{KJI}_{ABC}
 \label{tilde2}
\end{align}
for triple substitutions, and for quadruple substitutions:
\begingroup
\allowdisplaybreaks
\begin{align}
\tilde{E}^{IJKL}_{ABCD} =&
 \frac{258}{5040} \hat E^{IJKL}_{ABCD}
- \frac{48}{5040} \hat E^{IJLK}_{ABCD}
- \frac{48}{5040} \hat E^{IKJL}_{ABCD}
\nonumber\\&
- \frac{57}{5040} \hat E^{IKLJ}_{ABCD}
- \frac{57}{5040} \hat E^{ILIK}_{ABCD}
- \frac{48}{5040} \hat E^{ILKI}_{ABCD}
\nonumber\\&
- \frac{48}{5040} \hat E^{JIKL}_{ABCD}
+ \frac{78}{5040} \hat E^{JILK}_{ABCD}
- \frac{57}{5040} \hat E^{JLIK}_{ABCD}
\nonumber\\&
+ \frac{42}{5040} \hat E^{JLKI}_{ABCD}
+ \frac{42}{5040} \hat E^{JKIL}_{ABCD}
- \frac{57}{5040} \hat E^{JKLI}_{ABCD}
\nonumber\\&
- \frac{57}{5040} \hat E^{KIJL}_{ABCD}
+ \frac{42}{5040} \hat E^{KILJ}_{ABCD}
- \frac{48}{5040} \hat E^{KJIL}_{ABCD}
\nonumber\\&
- \frac{57}{5040} \hat E^{KJLI}_{ABCD}
+ \frac{78}{5040} \hat E^{KLIJ}_{ABCD}
+ \frac{42}{5040} \hat E^{KLJI}_{ABCD}
\nonumber\\&
+ \frac{42}{5040} \hat E^{LIJK}_{ABCD}
- \frac{57}{5040} \hat E^{LIKJ}_{ABCD}
- \frac{57}{5040} \hat E^{LJIK}_{ABCD}
\nonumber\\&
- \frac{48}{5040} \hat E^{LJKI}_{ABCD}
+ \frac{42}{5040} \hat E^{LKIJ}_{ABCD}
+ \frac{78}{5040} \hat E^{LKJI}_{ABCD}.
 \label{tilde3}
\end{align}
\endgroup

\subsection{Semi-biorthogonal formulation of residual equations}
\label{sec:BiorthHigherORdersKernel}
\noindent
Employing the semi-biorthogonal projections in Eqs.~\eqref{tilde2} and \eqref{tilde3}, the
CC equations Eqs.~\eqref{eq:CcEnergy1} and \eqref{eq:CcResidualEqFormal} are re-expressed as
\begin{align}
 E &= \braket{\Phi | \hat{H} \exp(\hat{T})|\Phi}_c,  \label{eq:CcEnergy} \\ 
r^{AB\cdots}_{IJ\cdots} &= \braket{\Phi|\tilde{E}^{IJ\cdots}_{AB\cdots} \hat{H} \exp(\hat{T})| \Phi}_c = 0.
 \label{eq:CcResidualEqFormal2}   
\end{align}

A peculiar looking feature of the so-modified equations is that, if are explicitly evaluated into tensor contractions,
they yield exclusively sets of permutation-related terms such as
\begin{align}
r^{ABC}_{IJK} = (\ldots) &+(-0.2)\;W^{AB}_{Ia} t^{Ca}_{JK}  \notag\\
                         &+(+1.0)\;W^{AB}_{Ia} t^{Ca}_{KJ}  \notag\\
                         &+(-0.2)\;W^{AB}_{Ja} t^{Ca}_{IK}  \notag\\
                         &+(-0.2)\;W^{AB}_{Ja} t^{Ca}_{KI}  \notag\\
                         &+(-0.2)\;W^{AB}_{Ka} t^{Ca}_{IJ}  \notag\\
                         &+(-0.2)\;W^{AB}_{Ka} t^{Ca}_{JI} + (\ldots).  \label{02}
\end{align}
in the triples case.
However the structure of such terms, which differ only by the permutation of lower indices $(IJK)$ appearing with prefactors $(1,-0.2,-0.2,-0.2,-0.2,-0.2)$, can be seen to reflect the zero eigenmode of the matrix $\mat M$ defined by the triple-substitution generalization of Eq.~\eqref{eq:MmatrixDoubles}: If written as a matrix eigenvalue problem
\begin{align}
\mat M \,\mat v = \lambda \mat v,
\end{align}
for the triples case, $\mat v=(1,1,1,1,1,1)^T$ is an eigenvector for $\lambda = 0$.
However, the existence of an eigenvector $\mat v$ with $\lambda=0$ implies that $\mat M\, \mat v = 0$; consequently, for the corresponding linear combinations of permutation-related de-excitation operators, \emph{no excited wave function can possibly have any overlap}.
Concretely, for the current case, if we define
\begin{align}
\bar{E}_{ABC}^{IJK} = \hat E_{ABC}^{IJK}
   +\hat E_{ABC}^{IKJ}
   +\hat E_{ABC}^{JIK}
   +\hat E_{ABC}^{JKI}
   +\hat E_{ABC}^{KIJ}
   +\hat E_{ABC}^{KJI}, \label{null}
\end{align}
---the linear combination of permutations corresponding to $\mat v=(1,1,1,1,1,1)^T$, then back-expanding $\mat M\, \mat v = 0$ translates into the relationships
\begin{align}
\langle \Phi | \bar{E}_{ABC}^{IJK} \hat{E}_{IJK}^{ABC} | \Phi \rangle &= 0\notag \\
\langle \Phi | \bar{E}_{ABC}^{IJK} \hat{E}_{IKJ}^{ABC} | \Phi \rangle &= 0\notag \\
\langle \Phi | \bar{E}_{ABC}^{IJK} \hat{E}_{JIK}^{ABC} | \Phi \rangle &= 0\notag  \\
\langle \Phi | \bar{E}_{ABC}^{IJK} \hat{E}_{JKI}^{ABC} | \Phi \rangle &= 0\notag  \\
\langle \Phi | \bar{E}_{ABC}^{IJK} \hat{E}_{KIJ}^{ABC} | \Phi \rangle &= 0\notag   \\
\langle \Phi | \bar{E}_{ABC}^{IJK} \hat{E}_{KJI}^{ABC} | \Phi \rangle &= 0,
\end{align}
for any indices $ABC$ and $IJK$.
As, furthermore, $\bra{\Phi}\bar{E}_{ABC}^{IJK}$ cannot possibly have any overlap with determinants $\hat E^{DEF\ldots}_{LMN\ldots}\ket{\Phi}$ in which either the total occupied-to-virtual excitation level differs from 3, or the set of indices $\{D,E,F\}$ differs from the set $\{A,B,C\}$, or the set $\{I,J,K\}$ differs from $\{L,M,N\}$, we can conclude that
\begin{align}
\langle \Phi | \bar{E}_{ABC}^{IJK} | \alpha \rangle = 0,\label{overlap2}
\end{align}
where $\ket{\alpha}$ is \emph{any} vector in the Fock space reachable by applying $\hat E$-operators to $\ket{\Phi}$.
As a consequence,
\begin{align}
   \braket{\Phi|\bar{E}_{ABC}^{IJK}\hat H\exp(\hat T)|\Phi}_c = 0,\label{overlap}
\end{align}
must identically vanish for $\bar{E}_{ABC}^{IJK}$ as defined in Eq.~\eqref{null}.
We note that Eq.~\eqref{overlap2} has been obtained in different circumstances before.\cite{kutzelnigg1985quantum,Helgaker:00r}
Results similar to Eq.~\eqref{overlap} can be obtained for higher order excitations, where then multiple linear combinations of residual permutations must vanish.

This Eq.~(\ref{overlap}) is, ultimately, what dictates the term structure in Eq.~\eqref{02} and the related terms:

For any ``real'' residual contribution (such as $X^{ABC}_{IJK}:= W^{AB}_{Ia} t^{Ca}_{KJ}$ in Eq.~\eqref{02}),
there are five additional residual contributions involving index permutations of $X^{ABC}_{IJK}$, of which the only role is accounting for the fact that the sum of all permuted residual contributions must satisfy Eq.~\eqref{overlap}.
It is the simple nature of this effect which allows us to take care of it without actually computing any of these additional permuted terms:
If, after computing the residuals $r^{ABC}_{IJK}$ we perform the a-posteriori transformation
\begin{align}
   {r}^{ABC}_{IJK} := r^{ABC}_{IJK}
   - \frac{1}{6}\big(r^{ABC}_{IJK}
   + r^{ABC}_{IKJ}
   + r^{ABC}_{JIK}
   + r^{ABC}_{JKI}
   + r^{ABC}_{KIJ}
   + r^{ABC}_{KJI}\big),\label{eq:TriplesResidualCleanup}
\end{align}
then even without the terms of Eq.~\eqref{02} with prefactor $-0.2$, the resulting residual $r^{ABC}_{IJK}$ will be compatible with Eq.~\eqref{overlap}.
\emph{We can therefore simply delete all such residual contributions and then restore their effect at runtime}---by performing a simple residual transformation such as Eq.~\eqref{eq:TriplesResidualCleanup} in the triples case.
For the present example of Eq.~\eqref{02}, we would explicitly compute only the residual contribution
\begin{align}
\tilde{r}^{ABC}_{IJK} &= (\ldots) +(1.2)\;W^{AB}_{Ia} t^{Ca}_{KJ} +  (\ldots) \label{03p}.
\end{align}
And then finally restore the effect of the missing terms (accounting for permutation relationships derived from the zero-eigenmodes of $\mat M$) by executing Eq.~\eqref{eq:TriplesResidualCleanup} after all triples residual contributions have been evaluated.

\begin{figure}
   \centering
   \setlength{\abovedisplayskip}{3pt}
   \setlength{\belowdisplayskip}{3pt}
   \setlength{\abovedisplayshortskip}{3pt}
   \setlength{\belowdisplayshortskip}{3pt}
      \rule{0.98\columnwidth}{\heavyrulewidth}
      \\[1.1ex]\includegraphics[width=.85\columnwidth]{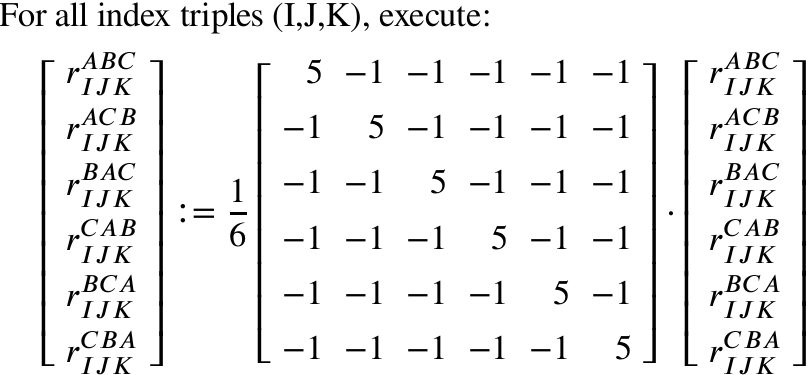}
      \rule{0.98\columnwidth}{\heavyrulewidth}
   \caption{Projective residual cleanup transformation for the CCSDT case. This linear transformation projects out all residual components in $r^{ABC}_{IJK}$ lying within the null space $\operatorname{ker}(\mat M)$ of the rank 3 permutation overlap matrix $\mat M$ (\emph{cf.} Eq.~\eqref{eq:MmatrixDoubles}).
   }
   \label{tab:PrjCleanup3}
\end{figure}

\begin{figure*}
   \centering
   \footnotesize
   \setlength{\abovedisplayskip}{3pt}
   \setlength{\belowdisplayskip}{3pt}
   \setlength{\abovedisplayshortskip}{3pt}
   \setlength{\belowdisplayshortskip}{3pt}
      \rule{0.94\textwidth}{\heavyrulewidth}
      \\[1.1ex]\includegraphics[width=.92\textwidth]{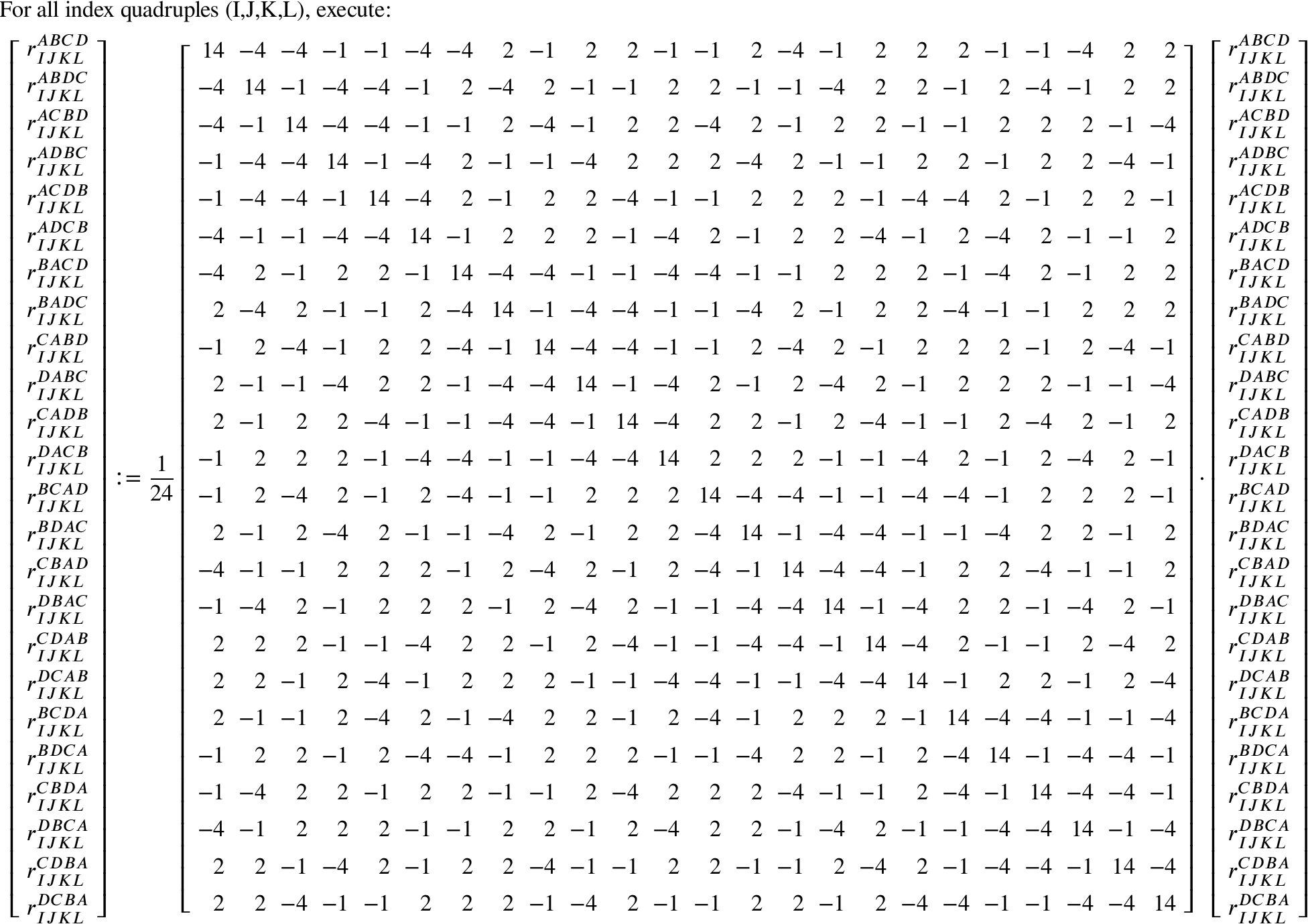}
      \rule{0.94\textwidth}{\heavyrulewidth}
   
   \caption{Projective residual cleanup transformation for the CCSDTQ case. This linear transformation projects out all residual components in $r^{ABCD}_{IJKL}$ lying within the null space $\operatorname{ker}(\mat M)$ of the rank 4 permutation overlap matrix $\mat M$ (\emph{cf.} Eq.~\eqref{eq:MmatrixDoubles}).}
   \label{tab:PrjCleanup4}
\end{figure*}

More generally, the residual cleanup transformation consists of projecting out all residual tensor components within the null-space of $\mat M$ (which, as explained, can never be reached if all equations are fully evaluated).
It can be written as a linear transformation of all permutation-related tensors; concrete realizations are given in Figs.~\ref{tab:PrjCleanup3} and \ref{tab:PrjCleanup4} for the triples and quadruples cases, respectively (the programs constructing those transformations are provided; see Sec.~\ref{sec:ProvidedPrograms}).
It is sufficient to perform this only once per iteration, after all raw residual contributions have been evaluated.
In the context of higher order CC methods, the residual cleanup transformation incurs a negligible computational cost.

As shown in Tab.~\ref{tab:NumTermsCcEquations}, with this additional deletion of terms, the use of the semi-bi-orthogonal operators leads to a massive reduction in the residual equation complexity (e.g., at CCSDTQ level, almost 95\% of the residual equations can be deleted).
The result of the here proposed scheme is a strongly reduced set of equations, which do not explicitly manifest redundancy relationships like Eq.~\eqref{overlap}, but which still give the correct coupled-cluster solution if the effect of these equations is re-established by a simple, cheap post-processing step at runtime, which has to be done only once per CC iteration.
For all practical purposes, this scheme can be regarded as a full extension of the biorthogonal projection scheme to higher than double substitutions, with all associated benefits.

\begin{table}[htbp]
  \caption{Number of unique tensor contractions contributing to the coupled-cluster residual equations defined with various projection schemes.
  ``Direct projection'' denotes the literal use of $r^{AB\cdots}_{IJ\cdots} := \bra{\Phi}\hat{E}^{IJ\cdots}_{AB\cdots}\ldots$  (Eq.~\eqref{eq:CcResidualEqFormal}) and ``Semi-Biorthogonal'' denotes the use of $r^{AB\cdots}_{IJ\cdots} := \bra{\Phi}\tilde{E}^{IJ\cdots}_{AB\cdots} \ldots$ (Eq.~\eqref{eq:CcResidualEqFormal2}) in conjunction with the $\tilde{E}^{IJ\cdots}_{AB\cdots}$ operators defined by Eqs.~\eqref{tilde2} and \eqref{tilde3}. In the latter case, ``(all Eqs.)'' denotes that all second-quantized terms arising in Eq.~\eqref{eq:CcResidualEqFormal2} are retained, and ``(non-redundant)'' denotes that all terms are deleted which only assure the fulfillment of permutation relations arising from $\mat M$'s zero eigenmodes (see text).
  }
    \begin{tabular}{lr@{\hspace*{1em}}r@{\hspace*{1em}}r}
    \toprule
          & CCSD  & CCSDT & CCSDTQ
    \\\midrule
    Direct projection &    225   &      3031 &  47709 \\
    Semi-Biorthogonal (all Eqs.)&    128$^a$   &     2901  & 47572 \\
    Semi-Biorthogonal (non-redundant) &       &    621   & 2534
    \\ \bottomrule
    \multicolumn{4}{l}{$^a$ For CCSD, the ordinary biorthogonal projection Eq.~\eqref{tilde} is used.}
    \end{tabular}%
  \label{tab:NumTermsCcEquations}%
\end{table}%

\section{Evaluation of matrix elements}\label{sec:EvaluatingMatrixElements}
In order to evaluate Eqs. 
(\ref{eq:CcEnergy}) and (\ref{eq:CcResidualEqFormal2})
in a practical computer program, they must be transformed into a series of tensor contractions involving amplitude, integral, and residual tensors.
This is commonly done using either diagrammatic expansions (e.g., Refs.~\onlinecite{Harris:99,Kallay:00}) or by employing operator-valued symbolic algebra and incrementally reducing binary products of substitution operators to sums of single substitution operators.\cite{Olsen:00,Hirata:00}
We here propose a scheme which can be viewed as a third alternative: The direct evaluation of coupling coefficients (DECC) with permutation group techniques.

Let us illustrate the DECC scheme with an example.
Consider the following contribution to the CCSD residual $r^{AB}_{IJ}$:
\begin{align}
      r^{AB}_{IJ} = \ldots + \frac{1}{2} \langle   \Phi | \hat E^{IJ}_{AB} \hat E^r_s \hat E^{ab}_{ij} | \Phi \rangle f^s_r t^{ij}_{ab} + \ldots. \label{eq:ExampleMatrixElement1}
\end{align}
The core of the scheme is the realization that, although the series of $\hat E$-operators leads to increasingly complex wave functions during their incremental application to the reference state ($\ket{\Phi}$, $\hat E^{ab}_{ij}\ket{\Phi}$, $\hat E^r_s \big(\hat E^{ab}_{ij}\ket{\Phi}\big)$, \ldots),
in total this term can only be non-zero if the lower indices $L=[A,B,s,i,j]$ are a permutation of the upper indices $U=[I,J,r,a,b]$.
Consequently, the matrix element must be representable in the general form
\begin{align}
      \langle \Phi | \hat E^{IJ}_{AB} \hat E^r_s \hat E^{ab}_{ij} | \Phi  \rangle = \sum_{\pi\in S_5} C(\pi) \delta^{U_1}_{L_{\pi(1)}}\delta^{U_2}_{L_{\pi(2)}}\ldots\delta^{U_5}_{L_{\pi(5)}},\label{eq:ExampleMatrixElement2}
\end{align}
where $\pi$ is a permutation, $S_N$ is the symmetric group of order $N$, and the prefactor $C(\pi)$ only depends on both the permutation $\pi$ and the form of the matrix element (but not on the concrete values of the indices in $U$ and $L$).

If we can evaluate $C(\pi)$ in closed form, then the full matrix element Eq.~\eqref{eq:ExampleMatrixElement1} can be obtained by simply iterating and summing over all involved index permutations multiplied with their associated $C(\pi)$ prefactors.

The coupling coefficient $C(\pi)$ can indeed be evaluated. A closer investigation (see appendix \ref{sec:EvaluatingTheCouplingCoefficient}) shows that
\begin{align}
   C(\pi) &= \mathcal{T}(\pi) \cdot (-1)^{n_\textrm{hole-co}} \cdot (-O)^{n_\textrm{cycles-in-$\pi$}}\label{eq:CcPrefactor}
\end{align}
where
\begin{itemize}
   \item $O$ is the occupancy of an occupied orbital (2 in the closed-shell spin 1/2 particle case considered here; for spin orbital substitutions it would be 1)
   \item $n_\textrm{cycles-in-$\pi$}$ is the number of cycles of which the permutation $\pi$ is composed. 
      [Note: A cycle of a permutation $\pi$ is a subset of points trading places with each other during incremental application of $\pi$. For example, the permutation $\pi=[4,5,1,3,2,6]$ has three cycles: (134), (25), and (6). (134) is a cycle since during an incremental application of $\pi\cdot\pi\cdot\pi\ldots$, all of the points 1,3,4 trade places only with each other, but not with 2,5 or 6. Permutation cycles are disjoint and can be trivially computed.]
   \item $n_\textrm{hole-co}$ is the number of ``hole contractions''---contractions in which the destruction operator stands to the right of the creation operator (\textit{i.e.}, the number of contractions involving occupied orbitals)
   \item $\mathcal{T}(\pi)$ is a topological factor taking account of the form of the matrix element and index domains: It is 1 or 0.
   It is 1 unless the permutation $\pi$ either (i) leads to an internal contraction within any $\hat E$-operator (e.g., aligns $r$ to $s$ in the example); (ii) implies the contraction of two indices $U_i$ and $L_{\pi(i)}$ to each other which lie in disjoint index domains (e.g., contracting an occupied orbital index $U_i$ to a virtual orbital index $L_{\pi(i)}$);
   (iii) results in a contraction in which either an upper occupied-orbital index $U_i$ stands to the right of its contracted lower index $L_{\pi(i)}$ (i.e., $U_i$ and $L_{\pi(i)}$ are occupied, but $i>\pi(i)$), or an upper virtual-orbital index $U_i$ stands to the left its contracted lower index $L_{\pi(i)}$ (i.e., $U_i$ and $L_{\pi(i)}$ are virtual, but $i<\pi(i)$). Note that neither case of (iii) can occur in example \eqref{eq:ExampleMatrixElement1}.
   Additionally, we here set $\mathcal{T}(\pi)$ to 0 for any permutations resulting in disconnected tensor contractions, since only connected contributions are needed in coupled cluster methods.
\end{itemize}
In total, we thus get a set of rules which closely resemble the full-contraction form of the generalized Wick theorem, \cite{kutzelnigg:GeneralizedNormalOrder} but without a restriction to reductions of binary operator products.

Consequently, contributions like Eq. \eqref{eq:ExampleMatrixElement1} can be evaluated by searching over the permutations of the lower indices, evaluating their prefactor by Eq.~\eqref{eq:CcPrefactor}, and substituting the resulting aligned indices into the tensor expressions. For example, the permutation $\pi=[4,5,2,1,3]$:
\begin{itemize}
   \item Aligns $U=[I,J,r,a,b]$ to $\pi(L)=[i,j,B,A,s]$.
   \item Has two hole contractions ($i$ to $I$ and $j$ to $J$) and two cycles ($[4,1]$ and $[3,5,2]$), and therefore a prefactor of $(-1)^2\cdot(-2)^2=4$
   \item And thus generates the residual contribution
      \begin{align}
            \frac{1}{2} (4\,\delta^I_i \delta^J_j \delta^r_B \delta^a_A \delta^b_s ) f^s_r t^{ij}_{ab} = 2 f^b_B t^{IJ}_{Ab}.
      \end{align}
\end{itemize}
The full set of coupled-cluster matrix elements in  
Eq.~\eqref{eq:CcEnergy} and Eq.~\eqref{eq:CcResidualEqFormal2}
can be computed by treating the individual sum terms of $\tilde E$ (Eqs.~\eqref{tilde}, \eqref{tilde0}, \eqref{tilde2}, \eqref{tilde3}), of $\hat H$ (Eq.~\eqref{Hop}), and of
$\exp(\hat T)=1 + \hat T + \frac{1}{2!}{\hat T}^2 + \frac{1}{3!}{\hat T}^3+\ldots$ with $\hat T=\hat T_1 + \hat T_2 + \hat T_3 +\ldots$ (Eq.~\eqref{eq:Top})
analogously.
For each sum term, the contributions resulting from lower index permutations are accumulated. This results in expressions which are flat tensor contractions and all coupling coefficients are resolved.

The outlined DECC scheme combines features of both the standard Wick theorem and diagrammatic techniques of evaluating second quantized matrix elements.
Like the Wick theorem techniques, it is fully algebraic and thus allows for computing matrix elements of almost any combinations of second quantized operators, without developing new rules of how to enumerate or weight the diagrams.
Like the diagrammatic techniques,\cite{Kallay:00} it allows for a simple computer implementation (including a direct enumeration of all involved tensor contractions) without invoking non-trivial operator algebra.
Of course, ultimately the three schemes are equivalent and lead to identical results, their only difference lying in interpretation and the complexity of implementation.

\section{Treatment of equivalent residual contributions}
\subsection{Invariances of tensor contraction expressions}\label{sec:InvariancesOfTensorExpressions}
The process described in Sec.~\ref{sec:EvaluatingMatrixElements} will generally lead to many residual contributions which are mathematically equivalent.
This can be a result of intrinsic index permutation symmetries of the involved tensors themselves, such as
\begin{align}
W_{ab}^{ij} &= W_{ba}^{ji} = W_{ib}^{aj} =  W_{bi}^{ja} = W_{aj}^{ib} = W_{ja}^{bi} = W_{ib}^{aj} = W_{bi}^{ja},  \label{s1}  \\
T_{ab}^{ij} &=  T_{ba}^{ji} \label{s2},
\end{align}
of the commutativity of number multiplication, leading to an invariance of tensor expressions regarding the ordering of their constituting tensor component terms, e.g.,
\begin{align}
T_{ab}^{ij} T_{cd}^{kl} = T_{cd}^{kl} T_{ab}^{ij} \label{s3},
\end{align}
of the invariance of tensor contraction expressions to the act of renaming the summation indices they involve, e.g., $(k \leftrightarrow l)$
\begin{align}
 W^{lI}_{kc} T^{lk}_{Ac} = W^{kI}_{lc} T^{kl}_{Ac} \label{s4},
\end{align}
of the invariance to renaming free indices in a tensor equation, as long as both sides of the equation are equally modified, e.g.
\begin{align}
                  \tilde{r}^{ABC}_{IJK} &\peq (1.2)\;W^{AB}_{Ia} t^{Ca}_{KJ}\notag
\\ \Longleftrightarrow\quad   \tilde{r}^{BAC}_{IJK} &\peq (1.2)\;W^{BA}_{Ia} t^{Ca}_{KJ} \label{s5},
\end{align}
and, finally, also combinations of all of these symmetries and invariances.

Despite the apparent mathematical simplicity of these transformations, efficiently handling them  in the general case is far from trivial in a computer program.
The reason for this is that the combination of all these invariance rules will in general lead to an factorial increase in the number of mathematically equivalent tensor expressions with an increase in the numbers of involved tensors or tensor indices.
For example, in an expression involving $n$ summation indices $p_1 p_2 \cdots p_n$, there are $n!$ ways of ordering them; if this is combined with the freedom to re-order tensors or apply intrinsic tensor permutational symmetries, one may quickly end up a very large number of terms, which are mathematically equivalent despite looking very different.
This can make it hard to decide whether or not two residual contributions generated by the second quantization algebra are equivalent or not.
In automated implementation techniques,\cite{janssen1991automated,Hirata:00,engelsputzka:factorization}
this issue is frequently ameliorated by grouping contractions according to topological properties, combined with various approaches to iterating over equivalent terms.

\subsection{Merging equivalent expressions via canonical forms}\label{sec:CanonicalForms}

In computer science, issues such as the above would typically be approached by a two-pass process:
First, one would decide on a \emph{canonical form} for a set of all equivalent objects (here tensor expressions---it does not matter what the canonical form is, it only matters that \emph{every single one} of the equivalent expressions gets mapped to the same one---the ``canonical representative'' of the set of equivalent expressions).
This transformation would then be applied to all generated residual contributions (at a cost of $\mathcal{O}(N_\mathrm{eq})$, where $N_\mathrm{eq}$ is the number of equations), and the canonicalized equations would then be sorted (at a cost of $\mathcal{O}(N_\mathrm{eq}\log(N_\mathrm{eq}))$), and their prefactors combined.

Unfortunately, the issue of deciding on a canonical form for a set of general tensor contraction expressions is itself non-trivial---at least if all of the invariances described in Sec.~\ref{sec:InvariancesOfTensorExpressions} are to be resolved.
However, the canonicalization problem has been addressed in a series of articles of Portugal and coworkers and Mart{\'\i}n-Garc{\'\i}a,\cite{portugal:DoubleCosetRep1,martingarcia:InvarPackage,martingarcia:xperm} who provided a practical algorithm to approach it---based on Butler's double-coset canonical representative algorithm\cite{butler:GroupHomomorphisms} from computational group theory.
The problem has been further investigated in a recent article by Li and coworkers, who provided algorithms with improved formal scaling for some of the group theoretical computations required in the canonicalization process,\cite{li2016classifications} and again by Niehoff who researched further algorithmic adjustments.\cite{niehoff:FasterTensorCanonicalization}
As far as we are aware of, this article by Li and coworkers,\cite{li2016classifications} which was developed independently of ours, is also the first published use of the double-coset technique and related computational group theory methods (such as stabilizer chains and the means to compute and use them) in the context of quantum chemistry.
However, for our program we developed and used a slight modification of Butler and Portugal's original approach, which we shall now describe.

\subsection{Adjustments to tensor expression and index notation}\label{sec:PermRepTensorNotation}
For illustration purposes, we will here rephrase tensor expressions involving upper and lower indices into flat expressions with only one kind of index, and explain the used transformations on the example of
\begin{align}
   r_{ABCD}^{IJKL} \peq W^{ab}_{ij} t_D^i t_a^L t_{AB}^{Ij} t_{Cb}^{KJ},
\end{align}
which is one of the residual contributions in CCSDTQ.
Formulated in flattened form, this becomes
\begin{align}
   \ti{r_4}{ABCDIJKL} \peq \ti{W}{abij}\!\!\ti{t_1}{Di}\!\!\ti{t_1}{aL}\!\!\ti{t_2}{ABIj}\!\!\ti{t_2}{CbKJ}. \label{eq:ExampleExpressions}
\end{align}
Additionally, \emph{free indices} (i.e., indices which appear on the lhs of an equation) will be written in upper case, while \emph{dummy indices} (i.e., indices which occur twice on the rhs of an equation and are implicitly summed over) will be written in lower case.
In this section, we will explicitly denote the rank of a residual or cluster amplitude tensor with a numeric index (e.g., $t_1$ is the tensor of single excitation amplitudes, $t_2$ of double excitation amplitudes, etc.), rather than inferring the rank from its number of indices.
We shall denote as \emph{slots} the places of a tensor expression into which symbolic indices can be inserted (for example, the 4-index integral tensor $W$ has four slots ($W[\sqcup\sqcup\sqcup\,\sqcup]$), and in ``$W[abij]$'' these slots are occupied by the indices $a,b,i,j$).

In the current term-by-term canonicalization,
we will further assume that the free indices of a residual contribution ($ABCDIJKL$ in Eq.~\eqref{eq:ExampleExpressions}) have already been brought into a canonical form on the lhs of the equation, as this is a straight-forward process.
For this reason, it is sufficient to treat only the expression on the rhs of such an equation, here
\begin{align}
   \ti{W}{abij} \ti{t_1}{Di} \ti{t_1}{aL} \ti{t_2}{ABIj} \ti{t_2}{CbKJ}, \label{eq:ExampleExpressionRhs}
\end{align}
while assuming that the free indices can no longer be renamed.

To closer reflect the actual implemented algorithms, additionally indexing will start at zero in this section, rather than one (i.e., an array of four elements will be indexed with $0,1,2,3$, and permutation indices start at 0, not at 1).

\subsection{Rephrasing canonicalization into group theory}\label{sec:CanonicalizationInGroupTheory}
The key discovery at the basis of the Butler-Portugal canonicalization approach for tensor expressions,\cite{butler:GroupHomomorphisms,portugal:DoubleCosetRep1,martingarcia:InvarPackage,martingarcia:xperm} such as Eq.~\eqref{eq:ExampleExpressionRhs},
is that the problem can be rephrased into the search of a canonical representative of a ``double coset''.
The double coset $\groupname{D} \groupelement{g} \groupname{S}$ is the set of permutations
\begin{align}
   \groupname{D} \groupelement{g} \groupname{S} := \left\{\groupelement{d}\cdot \groupelement{g}\cdot \groupelement{s}\;\big\vert\;\groupelement{d}\in\groupname{D},\,\groupelement{s}\in\groupname{S}    \right\},\label{eq:DoubleCoset}
\end{align}
where $\groupelement{g}\in S_{N}$ is a permutation of $N$ elements,
and both $\groupname{D}\subset S_N$ and $\groupname{S}\subset S_N$ are sub-groups of the symmetric group $S_N$ (the group of \emph{all} permutations of $N$ elements, not to be confused with $\groupname{S}$, which only contains slot permutations, see below).
The group action ``$\cdot$'' denotes the multiplication of permutations in the following convention:
\begin{align}
   \forall i\in\{0,1,\ldots,(N-1)\}:\;  (\groupelement{g}\cdot \groupelement{h})(i) := \groupelement{g}\big(\groupelement{h}(i)\big).\label{eq:PermutationGroupAction}
\end{align}
Note that both $\groupname{D}$ and $\groupname{S}$ are \emph{groups} (this means, in particular, that any combination of their respective elements will yield another group element---i.e., that they are ``closed under the group action'', and that for each element $\groupelement{x}$ included in the group, its inverse $\groupelement{x}^{-1}$ is also in the group), while the double-coset $\groupname{D} \groupelement{g} \groupname{S}$ is only a \emph{set}, and in general \emph{not} a group.
The problem of efficiently computing a canonical representative of a double coset (i.e., defining an algorithm which maps \emph{every single element} of a double coset $\groupname{D} \groupelement{g} \groupname{S}$ to the same single element a single element $\groupelement{g}'\in \groupname{D} \groupelement{g} \groupname{S}$) had originally been solved by Butler.\cite{butler:GroupHomomorphisms}

\begin{figure}
   \centering
   \includegraphics[width=.98\columnwidth]{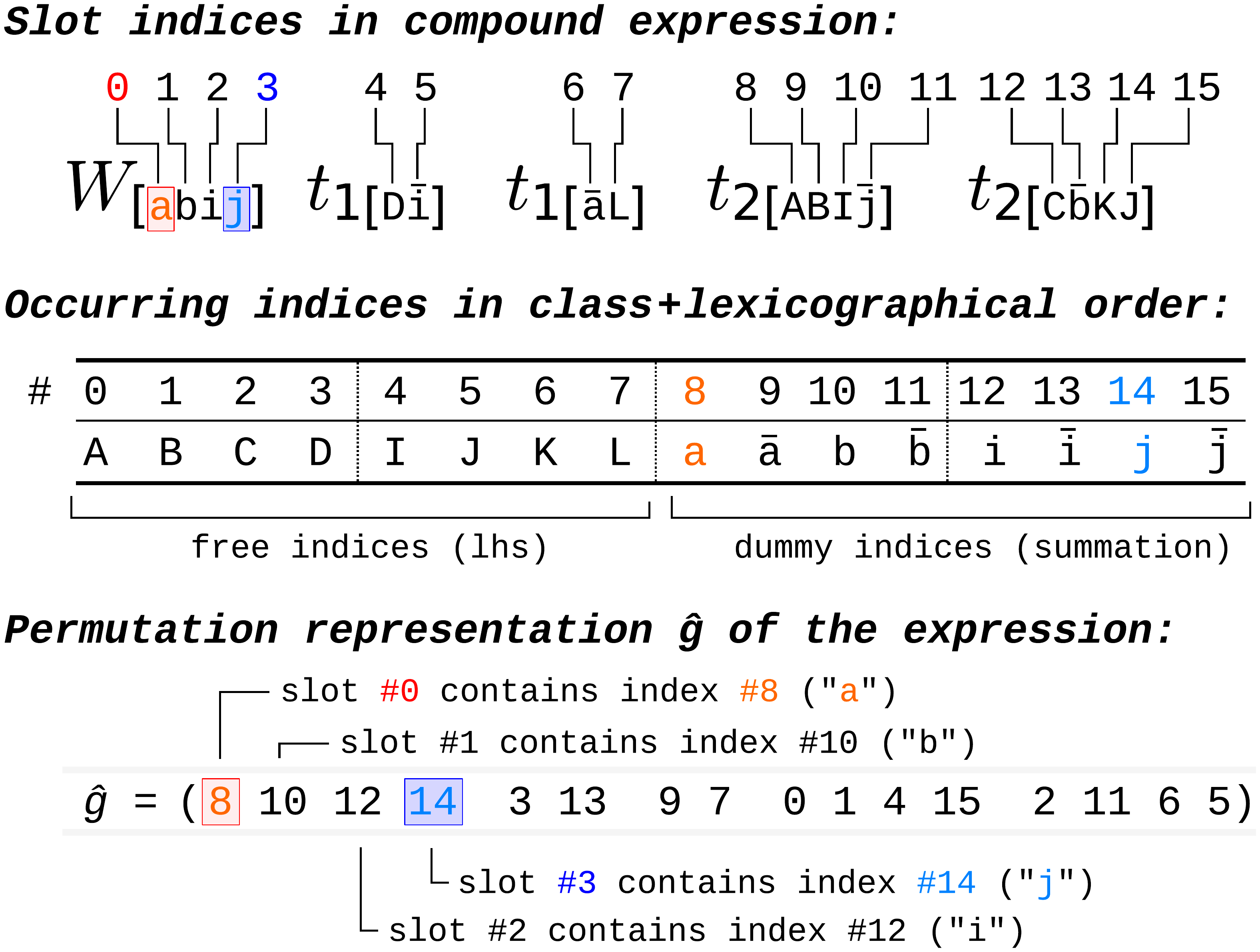}
   \caption{Example for the construction of the permutation representation $\hat g$ of Eq.~\eqref{eq:ExampleExpressionRhs}. See text for details. Some slots and indices have been highlighted in order to illustrate connections.}
   \label{fig:TensorPermRep}
\end{figure}

\begin{figure}
   \centering
   \includegraphics[width=.98\columnwidth]{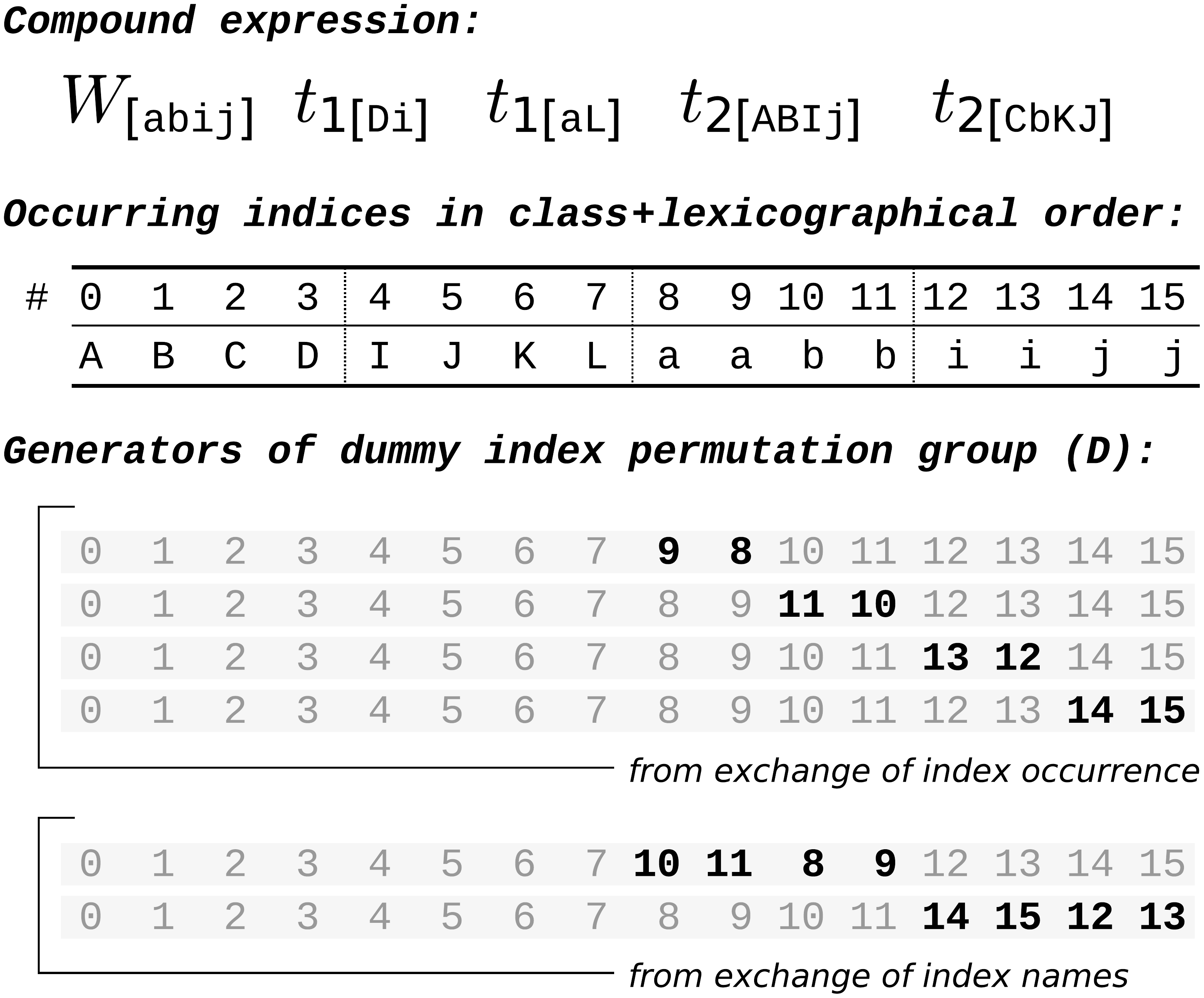}
   \caption{Construction of the dummy index symmetry group $\groupname{D}$ for the expression in Eq.~\eqref{eq:ExampleExpressionRhs}:
   All dummy indices occur twice, and these two occurrences can be exchanged (first four generators). Additionally, the names of all unique dummy indices \emph{within} a class of indices (e.g., all virtual dummy indices) can be exchanged with each other by index renaming (e.g., the last generator represents the exchange $i\leftrightarrow j$, meaning  $W^{ab}_{ij} t_D^i t_a^L t_{AB}^{Ij} t_{Cb}^{KJ}=W^{ab}_{ji} t_D^j t_a^L t_{AB}^{Ii} t_{Cb}^{KJ}$ for this concrete term)}
   \label{fig:TensorPermRepDummySymmetries}
\end{figure}

\begin{figure}
   \centering
   \includegraphics[width=.98\columnwidth]{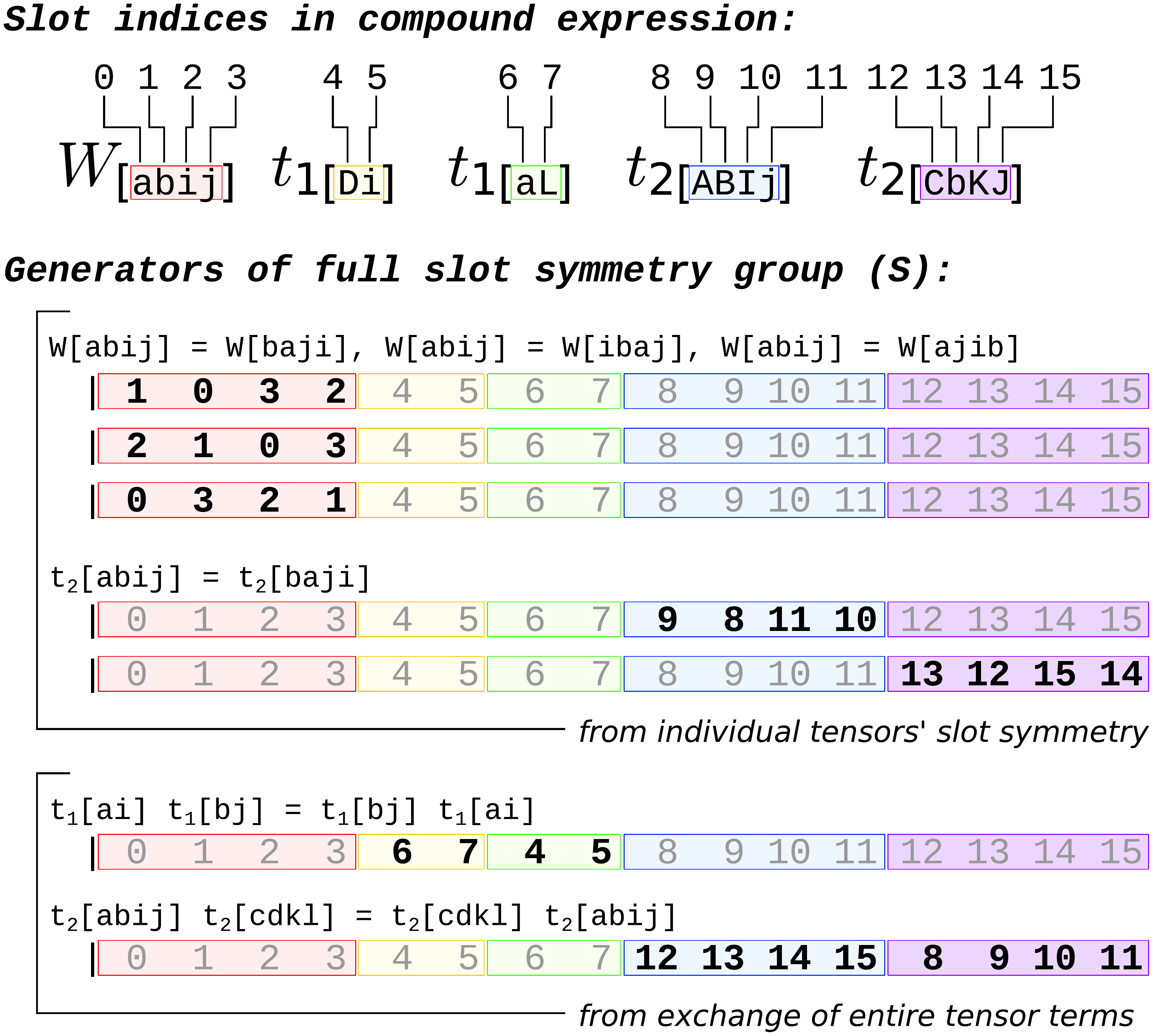}
   \caption{Construction of the slot symmetry group $\groupname{S}$ for the expression in Eq.~\eqref{eq:ExampleExpressionRhs}.
   }
   \label{fig:TensorPermRepSlotSymmetries}
\end{figure}

The translation of the tensor expression canonicalization problem into the double-coset canonical representative problem\cite{portugal:DoubleCosetRep1,martingarcia:InvarPackage,martingarcia:xperm} proceeds in three steps, starting with the definition of a tensor expression's permutation representation $\groupelement{g}$ as step one.
The significance of this representation is that all the expression transformations described in Sec.~\ref{sec:InvariancesOfTensorExpressions} will turn out to be directly representable by transformations
\begin{align}
   \groupelement{g}\quad\mapsto\quad \groupelement{g'} := \groupelement{d}\cdot\groupelement{g}\cdot\groupelement{s},
\end{align}
where $\groupelement{d}\in \groupname{D}$ and $\groupelement{s}\in \groupname{S}$ are elements of two permutation groups $\groupname{D}$ and $\groupname{S}$.
Steps two and three then are the concrete construction of the \emph{dummy label invariance transformation group} $\groupname{D}$ and the \emph{slot permutation symmetry group} $\groupname{S}$.
That is, the permutation representation provides us with a framework by which we can directly and uniquely parameterize the full set of tensor expression transformations which leave the expression's value invariant.
For the tensor expression in Eq.~\eqref{eq:ExampleExpressionRhs}, these three steps are illustrated in Figures \ref{fig:TensorPermRep}, \ref{fig:TensorPermRepSlotSymmetries}, and \ref{fig:TensorPermRepDummySymmetries}, respectively.

First, we define a permutation representation $\groupelement{g}\in S_N$ of the targeted tensor contraction expression (Fig.~\ref{fig:TensorPermRep}), where $N$ denotes the total number of its slots.
To this end, we first decide on a canonical order of the involved tensors themselves. We here define integral tensors ($W$, $f$) to preceed amplitude tensors ($t_1$, $t_2$, $\ldots$), and lower rank tensors to preceed higher rank tensors (i.e., $f<W$ and $t_1<t_2<t_3<\ldots$)---however, what exactly this order \emph{is} is insubstantial, as long as the same order is used for all expressions to be canonicalized together (even a lexicographical order based on only the tensor names would work, as long as each unique tensor has a unique name).
The tensors in the product expression are then brought into this order while preserving all their indices; for example, an expression 
\begin{align}
   \ti{t_1}{Di} \ti{t_2}{ABIj} \ti{t_1}{aL} \ti{W}{abij} \ti{t_2}{CbKJ}
\end{align}
would be reordered into
\begin{align}
   \ti{W}{abij} \ti{t_1}{Di} \ti{t_1}{aL} \ti{t_2}{ABIj} \ti{t_2}{CbKJ}.
\end{align}
For the ordered tensors, we then linearly index the $N$ \emph{slots} (cf. Sec.~\ref{sec:PermRepTensorNotation}) of the compound expression with the numbers $0,1,\ldots,(N-1)$ from left to right, as illustrated in Fig.~\ref{fig:TensorPermRep} (top).
We then collect all index labels placed in the slots of the compound expression, in original order, from left to right, and store them as the list $I_o$ (here: $I_o=[a,b,i,j,D,\bar i,\bar a,L,A,B,I,\bar j,C,\bar b,K,J]$).
Note that each dummy label occurs \emph{twice}; to clarify this aspect, we here explicitly denote the second occurrence of a dummy label with a bar (i.e., $a$ is contracted to $\bar{a}$, but an exchange of $a$ and $\bar{a}$ in the compound expression is inconsequential).
Further, we define as $I_c$ the ``canonical'' list of index labels obtained by sorting the labels in $I_o$ first by class (free indices preceed dummy indices, and virtual indices preceed occupied indices) and then lexicographically by name (here: $I_c=[A,B,C,D,I,J,K,L,a,\bar a,b,\bar b,i,\bar i,j,\bar j]$; see Fig.~\ref{fig:TensorPermRep}, middle).
With this identification of the ordered slot indices in hand, as well as the label lists $I_o$ and $I_c$, we are ready to define the permutation representation $\groupelement{g}\in S_N$ of the compound tensor expression:
For each slot index $i\in\{0,1,\ldots,(N-1)\}$, the number $\groupelement{g}(i)$ denotes which element of the sorted label list $I_c$ is placed in the $i$'th slot of the compound expression (Fig.~\ref{fig:TensorPermRep}, bottom). Or, rephrased, $\groupelement{g}$ is the permutation with the property
\begin{align}
   \forall i\in\{0,1,\ldots,(N-1)\}:\;I_o[i] = I_c[\groupelement{g}(i)],\label{eq:TensorPermRepIndices}
\end{align}
where $I[i]$ denotes the $i$'th element of a list $I$.

Second, we identify $\groupname{D}$, the group of (dummy) index label transformations leaving the tensor expression's value invariant (Fig.~\ref{fig:TensorPermRepDummySymmetries}).
To this end, first note that with the permutation representation $\hat g$ as defined above, the actions of either (i) exchanging the names of two pairs of dummy indices of a common one-particle space (e.g., replacing $i,\bar{i}$ by $j,\bar{j}$ and vice-versa), or (ii) exchanging the slots to which the two occurrences of a single pair of dummy labels are assigned (e.g., replacing $W[abij]\;t[D\bar{i}]$ by $W[ab\bar{i}j]\;t[Di]$), can both be represented by 
\begin{align}
   \groupelement{g}\quad\mapsto\quad \groupelement{g'}:=\groupelement{d}\cdot \groupelement{g}.
\end{align}
That is, by applying a permutation $\groupelement{d}$ to the \emph{image} of $\groupelement{g}$.
This is seen when combining the definition of the group action (Eq.~\eqref{eq:PermutationGroupAction}) with the definition of the tensor expression permutation representation (Eq.~\eqref{eq:TensorPermRepIndices}); together, they imply that 
$\groupelement{g'}=\groupelement{d}\cdot \groupelement{g}$ represents the tensor expression with the index list
\begin{align}
   I'[i] = I_c[\groupelement{g}'(i)] = I_c[\groupelement{d}\big(\groupelement{g}(i)\big)].\label{eq:PermRepIndexList}
\end{align}
Effectively, this means that left-applying $\groupelement{d}$ to $\groupelement{g}$ has the same effect as applying the permutation $\groupelement{d}^{-1}$ to the sorted index list $I_c$ while retaining the original $\groupelement{g}$. The permutation $\groupelement{d}$ can be viewed as acting on the sorted index list $I_c$!
So if we define as $\groupname{D}$ the group generated by all the permutations accounting for the cases (i) and (ii) above (note that this encompasses only exchanges within either two or four consecutive elements of a permutation to obtain a full set of generators, see Fig.~\ref{fig:TensorPermRepDummySymmetries}), then any $\groupelement{g}'\in\{\groupelement{d}\cdot \groupelement{g};\;\groupelement{d}\in\groupname{D}\}$ represents a tensor expression which is mathematically equivalent to the one represented by $\groupelement{g}$ itself.

Third, we identify the group $\groupname{S}$ of admissable slot permutations of the compound expression (Fig.~\ref{fig:TensorPermRepSlotSymmetries}).
To this end, first note that a transformed permutation
\begin{align}
   \groupelement{g}\quad\mapsto\quad \groupelement{g'}:=\groupelement{g}\cdot\groupelement{s}
\end{align}
corresponds to a tensor expression with the index list
\begin{align}
   I'[i] = I_c[\groupelement{g}'(i)] = I_c[\groupelement{g}\big(\groupelement{s}(i)\big)].
\end{align}
That is, right-multiplication of a permutation $\groupelement{s}$ to $\groupelement{g}$ 
corresponds to a permutation of the \emph{slots} of the compound expression
to which the index labels in $I_c$ are assigned.
As illustrated in Fig.~\ref{fig:TensorPermRepSlotSymmetries}, there are two distinct mechanisms by which a permutation of the slots-to-label assignment of the compound expression may yield a mathematically equivalent tensor expression:
(i) intrinsic slot-permutation symmetries of the \emph{individual} tensors occurring in the tensor expression (this typically reflects some physical properties of the involved tensors), and (ii) if a unique source tensor appears in multiple different instances in the tensor expression (such as $t_1$ appearing two times in Fig.~\ref{fig:TensorPermRepSlotSymmetries}), then their full set of slots may be exchanged.
If we define as $\groupname{S}$ the group of slot permutations in the compound expression which is spanned by all generators described in (i) and (ii), then any transformed $\groupelement{g'}\in\{\groupelement{g}\cdot\groupelement{s};\;\groupelement{s}\in \groupname{S}\}$ will represent a tensor expression mathematically equivalent to $\groupelement{g}$.

As the dummy label permutations described by $\groupname{D}$ and the slot permutations described by $\groupname{S}$ do not interfere with each other and can be applied independently, we conclude that all the elements $\groupelement{g}'\in \groupname{D}\groupelement{g}\groupname{S}$ of the double coset defined in Eq.~\eqref{eq:DoubleCoset} are permutation representations of mathematically equivalent tensor expressions.
Moreover, as all invariances described in Sec.~\ref{sec:InvariancesOfTensorExpressions} are covered by these transformations, the double coset $\groupname{D}\groupelement{g}\groupname{S}$ contains the permutation representations of \emph{all} equivalent tensor expressions.
So by constructing $\groupname{D}$, $\groupelement{g}$, and $\groupname{S}$ for a given input tensor expression, then finding the double coset's canonical representative $\groupelement{g}'_\mathrm{can}\in\groupname{D}\groupelement{g}\groupname{S}$, and then re-constructing the tensor expression represented by this $\groupelement{g}'_\mathrm{can}$ via Eq.~\eqref{eq:TensorPermRepIndices}, any input tensor expression can be transformed into a canonical form, solving the original problem.
The only question left open is how to define and compute the canonical representative $\groupelement{g}'_\mathrm{can}$ of the double coset $\groupname{D}\groupelement{g}\groupname{S}$, given an input $\groupelement{g}$ and the sets of permutations generating $\groupname{D}$ and $\groupname{S}$.

\subsection{Finding the double-coset canonical representative}
As explained in Sec.~\ref{sec:CanonicalizationInGroupTheory}, the tensor product canonicalization problem can be rephrased into the group theoretical problem of defining a function
$\canfn(\groupelement{g'})$, which for each permutation $\groupelement{g'}$ in the double coset $\groupname{D}\groupelement{g}\groupname{S}$ returns the same unique canonical permutation $\groupelement{g}_\mathrm{can}\in\groupname{D}\groupelement{g}\groupname{S}$:
\begin{align}
   \forall \groupelement{g'} \in \groupname{D}\groupelement{g}\groupname{S}:\; \canfn(\groupelement{g'}) = \groupelement{g}_\mathrm{can} \in \groupname{D}\groupelement{g}\groupname{S}.\label{eq:DefCanFn1}
\end{align}
In this, $\groupelement{g}$ is a permutation, and $\groupname{D}$ and $\groupname{S}$ are two permutation groups (given in terms of their respective generators);  $\groupelement{g}$, $\groupname{D}$ and $\groupname{S}$ are all determined from the input tensor expression.
$\groupelement{g}_\mathrm{can}$ is called the ``double coset canonical representative''; it depends on $\groupname{D}$, $\groupname{S}$, and $\groupname{g}$, but not on the input $\groupelement{g'} \in \groupname{D}\groupelement{g}\groupname{S}$.
As long as these conditions are fulfilled, how exactly the function $\canfn(\groupelement{g'})$ is defined is inconsequential for the canonicalization process.

\begin{figure}
   \centering
   \rule{0.98\columnwidth}{2pt}
   \\[1.1ex]\includegraphics[width=0.86\columnwidth]{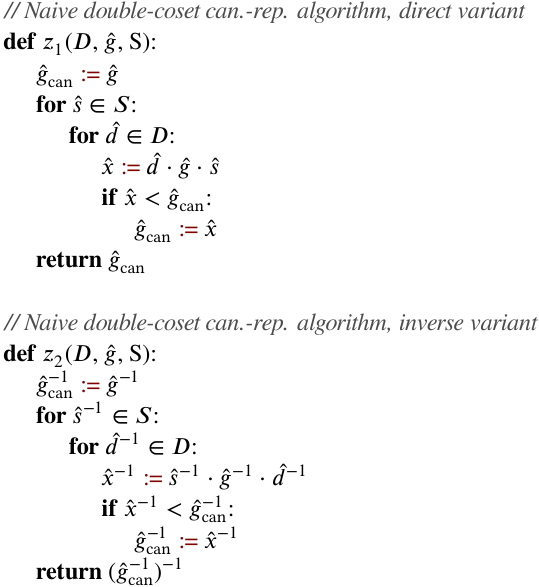}
   \rule{0.98\columnwidth}{2pt}
   \caption{Pseudo-code for two naive algorithms for computing the canonical representative of a double-coset $\groupname{D}\groupelement{g}\groupelement{S}$. See text.
   }
   \label{fig:DoubleCosetCanRepNaive}
\end{figure}

These requirements directly suggest two straight-forward algorithms for defining the canonicalization function, which are given in Fig.~\ref{fig:DoubleCosetCanRepNaive}.
In the first ``naive'' variant, $\canfn_1(\groupelement{g})$, we directly iterate over all group elements $\groupelement{s}\in\groupname{S}$ and $\groupelement{d}\in\groupname{D}$, and for each of those explicitly evaluate the double coset element $\groupelement{x}:=\groupelement{s}\cdot \groupelement{g}\cdot \groupelement{d}$. The algorithm then simply returns the minimal such $\groupelement{x}$ under the lexicographical order of permutations.
This algorithm does solve the canonicalization problem, and, in particular, can straightforwardly reduce the computational scaling of the merging of equivalent equations to $\mathcal{O}(N_\mathrm{eq}\log(N_\mathrm{eq}))$ in the number of equations $N_\mathrm{eq}$ as explained in Sec.~\ref{sec:CanonicalForms}. However, the computational cost of this method per individual equation can be high, as both $\groupname{S}$ and $\groupname{D}$ can contain \emph{many} elements.
Towards reducing this cost, we will now discuss a number of variations of the algorithm.

As first step towards this goal, consider the second variant of the naive algorithm in Fig.~\ref{fig:DoubleCosetCanRepNaive}, $\canfn_2(\groupelement{g})$.
This second variant $\canfn_2(\groupelement{g})$ differs from the first by iterating over the inverses of the double coset elements $(\groupelement{s}\cdot \groupelement{g}\cdot \groupelement{d})^{-1}=\groupelement{d}^{-1}\cdot \groupelement{g}^{-1}\cdot \groupelement{s}^{-1}$, rather than the double coset elements directly.
In general, $\canfn_2$ produces a different canonical representative than the algorithm $\canfn_1$ (because now inverse permutations are lexicographically compared), but as explained before, this is inconsequential as long as the same canonical representative is produced for any input permutation in the double coset $\groupname{D}\groupelement{g}\groupelement{S}$.
While $\canfn_2$ may look algorithmically more complex than $\canfn_1$, this is not really the case:
Note that as $\groupname{S}$ is a group, $\groupelement{s}\in\groupname{S}$ implies $\groupelement{s}^{-1}\in\groupname{S}$; so iterating over all $\groupelement{s}\in\groupname{S}$ or over all $\groupelement{s}^{-1}\in\groupname{S}$ are actually identical operations. The same applies for group $\groupname{D}$.
Note also that in the second variant, the $(\cdot)^{-1}$ in $\groupelement{g}_\mathrm{can}^{-1}$ and $\groupelement{x}^{-1}$ can be regarded as part of the name of the corresponding objects, rather than as an actual inversion operation (i.e., one would directly store the inverse permutation $\groupelement{x}^{-1}$, rather than $\groupelement{x}$; the only two actual inversion operations in the entire procedure would happen in the first step where $\groupelement{g}^{-1}$ is assigned to $\groupelement{g}_\mathrm{can}^{-1}$ and in the last step where $(\groupelement{g}_\mathrm{can}^{-1})^{-1}$ is returned).

\begin{figure}
   \centering
   \rule{0.98\columnwidth}{2pt}
   \\[1.1ex]\includegraphics[width=0.93\columnwidth]{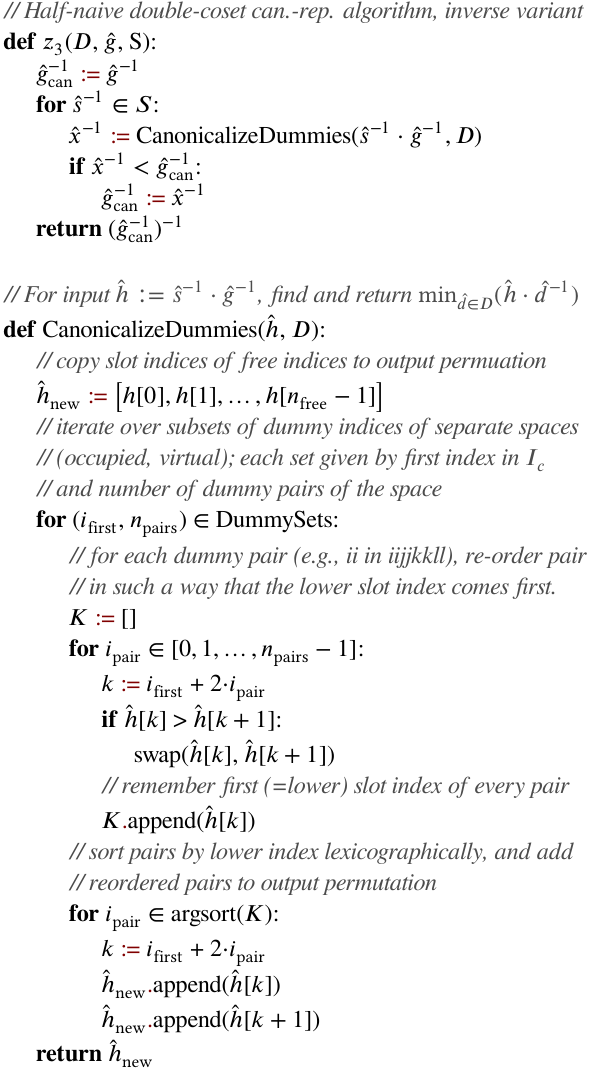}
   \rule{0.98\columnwidth}{2pt}
   \caption{Pseudo-code for the half-naive algorithm $z_3(\groupelement{g})$ of computing the canonical representative of a double-coset $\groupname{D}\groupelement{g}\groupelement{S}$.
   The integer $n_\mathrm{free}$ denotes the number of free indices (i.e., non-summation) indices in the tensor expressions; these are always the first $n_\mathrm{free}$ indices in the canonical index list $I_c$ (for the example in Fig.~\ref{fig:TensorPermRep}, we would have $n_\mathrm{free}=8$ for the first eight indices $ABCDIJKL$ in $I_c$).
   The permutable sets of dummy indices in each index domain (occupied, virtual) are given by tuples $(i_\mathrm{first}, n_\mathrm{pairs})$; 
   For the example in Fig.~\ref{fig:TensorPermRep}, these would be $(i_\mathrm{first}=8, n_\mathrm{pairs}=2)$ for the two virtual dummy pairs $a\bar{a}$ and $b\bar{b}$ and 
   $(i_\mathrm{first}=12, n_\mathrm{pairs}=2)$ for the two occupied dummy pairs $i\bar i$ and $j\bar j$.
   This algorithm produces identical canonical representatives as $z_2(\groupelement{g})$ (Fig.~\ref{fig:DoubleCosetCanRepNaive}), but saves the potentially expensive operation of iterating over the permutations in $\groupname{D}$.
   }
   \label{fig:DoubleCosetCanRepHalfNaive}
\end{figure}

In the inverse form $\canfn_2(\groupelement{g})$ of the naive algorithm, all the core operations are performed on permutations $\groupelement{g}^{-1}$ which map from indices into the canonical label list $I_c$ to the slot index in which a given label of $I_c$ stands; so $(\groupelement{g}^{-1})[i]$ denotes the slot index into which the canonical index label $I_c[i]$ is mapped, rather than the other way around as defined in Eq.~\eqref{eq:PermRepIndexList}.
The core point of this inverse reformulation of the problem is this: If combined with the near-trivial form of the dummy permutation group $\groupname{D}$ (see Fig.~\ref{fig:TensorPermRepDummySymmetries}), it allows removing the iterations over $\groupelement{d}^{-1}\in \groupname{D}$ in the naive algorithm, because this lexicographical minimization over $\groupname{D}$ can be easily done explicitly by combining simple index sorting operations.
These index sorting operations come at a computational cost of at most $\mathcal{O}(N_d \log(N_d))$ where $N_d$ is the number of dummy index labels, rather than $\mathcal{O}(N_d!)$ in the worst case if $D$ is explicitly iterated over.
This yields the half-naive algorithm $z_3(\groupelement{g})$ presented in Fig.~\ref{fig:DoubleCosetCanRepHalfNaive}.

This leaves as potentially problematic part only the explicit iteration over the group elements $\groupelement{s}^{-1}\in\groupname{S}$.
This minimization, too, can be transformed into a computationally efficient form.
However, a detailed description of the required algorithm requires the stabilizer chain representation of a permutation group of the field of computational group theory,\cite{hulpke:NotesOnComputationalGroupTheory,seress:PermutationGroupAlgorithms} which we cannot describe in detail here, but rather just broadly outline the core idea:
It can be shown\cite{hulpke:NotesOnComputationalGroupTheory,seress:PermutationGroupAlgorithms} that for each permutation group $\groupname{P}$ of $N$ integers $\{0,1,\ldots,(N-1)\}$, it is possible to obtain a list of points $(b_0, b_1, \ldots, b_{B-1})$,  $b_i\in\{0,1,\ldots,(N-1)\}$ called a ``base'', such that
\begin{align}
   \groupname{P} \supset P_{b_0} \supset \groupname{P}_{(b_0,b_1)} \supset \ldots \supset \groupname{P}_{(b_0,b_1,\ldots,b_{B-1})} = \{\groupelement{1}_N\},\label{eq:StabilizerChain}
\end{align}
where $\{\groupelement{1}_N\}$ denotes the (trivial) permutation group consisting of only the identity permutation of length $N$. The sub-group
\begin{align}
   \groupname{P}_{(b_0,b_1,\ldots)} := \{\groupelement{p}\in\groupname{P}\;|\; p(b_0)=b_0 \land p(b_1)=b_{1} \land \ldots \}
\end{align}
denotes the \emph{stabilizer} of the set of points $b_0,b_1,\ldots$;
that is, the set of all permutations $\groupelement{p}\in\groupname{P}$ which have the property that all the selected points $b_0, b_1,\ldots$ are invariant under $\groupelement{p}$: $\forall i\in\{0,1,\ldots\}:\;\groupelement{p}(b_i) = b_i$.
If $\groupname{P}$ is a group, the stabilizer of a set of points is obviously also a group.
For a given permutation group $\groupname{P}$, its stabilizer chain representation Eq.~\eqref{eq:StabilizerChain} therefore affords a decomposition into a nested set of simpler and simpler sub-groups, each of which has the property of stabilizing an additional base point $b_k$.
While the algorithm is not explicitly given here, one can imagine that by invoking this stabilizer chain representation for the slot symmetry group $\groupname{S}$, it is then possible to replace the direct iteration over $\groupelement{s}^{-1}\in\groupname{S}$ in Fig.~\ref{fig:DoubleCosetCanRepHalfNaive} by incremental minimizations over the elements of the cosets of the stabilizer chain, each of which individually contains only a small number of elements.
The core elements of this reformulation are first constructing a base which is itself ordered, and then re-ordering the tensor slots such that the (ordered) base points of the stabilizer chain of $\groupname{S}$ come first, followed by all other slot indices.
For the detailed double coset canonicalization algorithm, we refer to the documented source code of the provided example programs (see Sec.~\ref{sec:ProvidedPrograms}), and we refer to the textbooks of computational group theory\cite{hulpke:NotesOnComputationalGroupTheory,seress:PermutationGroupAlgorithms} for the computation and use of stabilizer chains.

The stabilizer chain representation also allows for efficient means of representing permutation groups given by generators, and for iterating over the group elements if required.
In our current program, we used the \emph{PermutationGroup} class of SymPy\cite{bib:Sympy} and its implementation of the Schreier-Sims algorithm and of base swaps to obtain the stabilizer chain representations with ordered base and corresponding strong generators.

\section{Provided Programs}\label{sec:ProvidedPrograms}
\begin{figure}
   \centering
   \rule{0.98\columnwidth}{2pt}
   \\[1.1ex]\includegraphics[width=0.96\columnwidth]{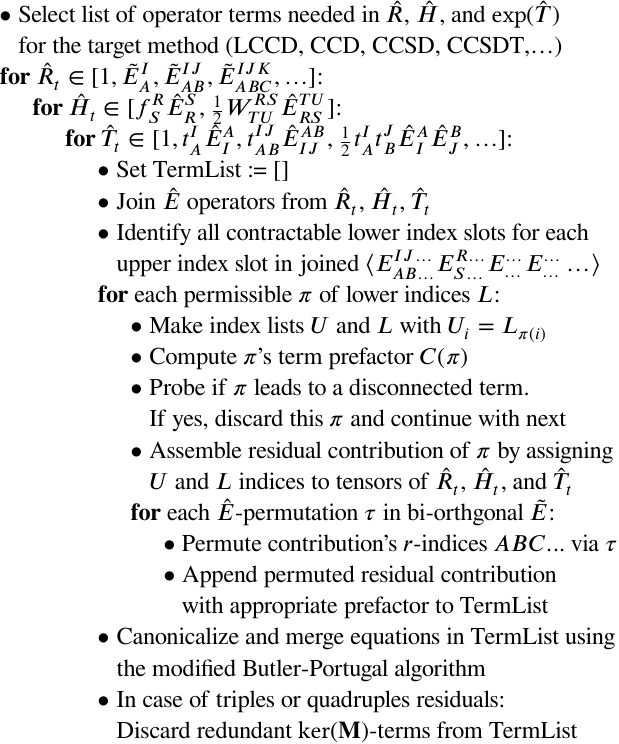}
   \rule{0.98\columnwidth}{2pt}
   \caption{Pseudo-code for the algorithm used to assemble the CC working equations with the DECC scheme as used in the provided program \emph{cc-eqs-cs}.}
   \label{fig:GeneratingTheWorkingEquations}
\end{figure}

The described techniques have been implemented into a Python program, called \emph{cc-eqs-cs}, for deriving and simplifying the closed-shell coupled-cluster equations up to CCSDTQ using the here described permutation group techniques.
This program combines all aspects described in the theory sections: fully spin-free excitation operators, normal-ordered Hamiltonians, linear combination of contravariant projection, elimination of the redundant terms, and canonicalization of the symmetry equivalent terms. The concrete algorithm is summarized in Figure~\ref{fig:GeneratingTheWorkingEquations}.
The output of the program is an equation list, complemented by generated C++ code encoding these equation lists as data elements.

The \emph{cc-eqs-cs} program is accompanied by a separate Python script \emph{make\_covariant\_proj.py} which implements the derivation of the semi-biorthogonal de-excitation operators $\tilde{E}^{IJK}_{ABC}$ in Eq.~\eqref{tilde2} and $\tilde{E}^{IJKL}_{ABCD}$ in Eq.~\eqref{tilde3}, and the projective residual cleanup transformations in Figs.~\ref{tab:PrjCleanup3} and \ref{tab:PrjCleanup4}.

\begin{figure}
   \centering
   \rule{0.98\columnwidth}{2pt}
   \\[1.1ex]\includegraphics[width=0.96\columnwidth]{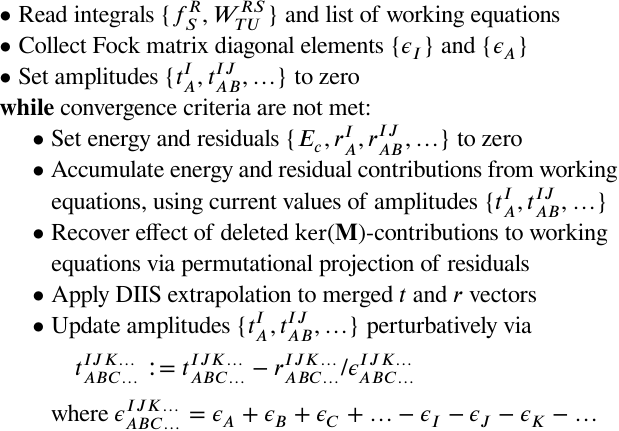}
   \rule{0.98\columnwidth}{2pt}
   \caption{Pseudo-code for iteratively computing CC wave functions with the here-proposed semi-biorthogonal residual projections: In each iteration, only the reduced set of working equations is explicitly evaluated, from which all $\operatorname{ker}(\mathbf{M})$ contributions have been deleted as described in Sec.~\ref{sec:BiorthHigherORders}.
   The missing equation's effect is then restored a-posteriorly by a cheap permutational projection of the triples and quadruples residual tensors $r^{IJK}_{ABC}$ and $r^{IJKL}_{ABCD}$ (Sec.~\ref{sec:BiorthHigherORders}).
   This approach is used in the provided program \emph{srci}.}
   \label{fig:EvaluatingTheWorkingEquations}
\end{figure}
The working equation lists are processed by a prototype C++ program, called \emph{srci}, which iteratively solves the CC (and CI) equations based on integral data files.
The algorithm to do so is summarized in Fig.~\ref{fig:GeneratingTheWorkingEquations}; it follows the standard paradigm to iteratively solve CC equations, apart from the inserted residual cleanup step discussed in Sec.~\ref{sec:BiorthHigherORders} (see Fig.~\ref{fig:GeneratingTheWorkingEquations}), and evaluating the CC energy by a generalization of the Hylleraas functional rather than Eq.~\eqref{eq:CcEnergy1} directly, which slightly improves iterative energy convergence.
In order to evaluate the residual tensors, the program invokes a very general tensor contraction kernel which is capable of evaluating generic tensor contraction expressions (including contractions involving more than two source terms) using an automated term-by-term factorization which yields a correct scaling in terms of computational resources.
\emph{srci} is a prototype program---it is flexible and simple, but not efficient or fast.

These programs are freely available, including source code, on the homepage of the Knizia group (http://sites.psu.edu/knizia/).
Due to their simplicity and straight-forwardness, the programs also offer considerable flexibility for testing and implementing other electronic structure methods.

The correctness of the described approach to the CC equations, of the generated equations lists, and the programs implementing them, has been established by comparison of numerical results to Kall\'{a}y's general order coupled-cluster program \emph{MRCC}\cite{kallay2011mrcc} and an in-house full-CI program \emph{fci}, which is also available on the Knizia group homepage.
Test Hamiltonians have been generated as {FCIDUMP}s with Molpro's determinant-FCI program. \cite{werner2012molpro,knowles1984new,knowles1989determinant}

\section{Summary, Conclusions, and Outlook}

We presented a general-order spin-free formulation of the single-reference closed-shell coupled-cluster method, which is based on two contributions:
(1) we proposed an efficient and simple way of dealing with the bi-orthogonal contravariant projection problem for higher orders of coupled-cluster; and (2)
we proposed a way to derive and simplify the required matrix elements via permutation group and double coset techniques.

Contribution (1) may add an important ingredient to the construction of highly efficient closed-shell CCSDT and CCSDTQ programs, which are frequently used in high-accuracy thermochemistry\cite{karton2006w4,Stanton:04}: the complete elimination of spin degrees of freedom, as well as the simple permutational structure of the amplitude and residual tensors (compared to a spin-orbital formulation) have potential to reduce the computational cost of the higher-order CC method by a large prefactor---without any approximation.
Contribution (2) provides a theoretically simple scheme for computing almost any second-quantized matrix elements, which furthermore affords using spin-free operators directly, and is highly suitable for a computer-based equation derivation.
We therefore expect this technique to be useful in the construction and testing of novel variants of coupled-cluster theory as well as other novel electronic structure methods based on second quantization.

Apart from the use in equation canonicalization invoked here, we believe that the permutation representation of tensor contractions and the used permutation group techniques of computational group theory may also become useful in other contexts in quantum chemistry---not only in the generation and simplification of working equations, but also in their efficient evaluation with generic tensor contraction kernels.
We intend to explore these possibilities in future research.

\bigskip

\noindent{\bf Supporting Information}\\
Full equation lists for the semi-biorthogonal closed-shell CCSD, CCSDT, CCSDTQ residuals. Source codes of programs for deriving the equations (\emph{cc-eqs-cs.py}), for constructing the semi-biorthogonal projections (\emph{make\_covariant\_proj.py}), and for numerically evaluating the working equations for a given Hamiltonian (\emph{srci}).

\bigskip

\noindent{\bf Funding}\\
Parts of this project were supported by a startup fund from the Pennsylvania State University.

\appendix

\section{Comments on Eq. (\ref{eq:ExampleMatrixElement2})}\label{sec:EvaluatingTheCouplingCoefficient}

We here explain the reasoning behind Eqs.~\eqref{eq:ExampleMatrixElement2} and \eqref{eq:CcPrefactor}, by illustrating their emergence for the example from Eq.~\eqref{eq:ExampleMatrixElement1}:
\begin{align}
   \braket{\Phi | \hat E^{IJ}_{AB} \hat E^r_s \hat E^{ab}_{ij} | \Phi} = \sum_{\pi\in S_5} \mathcal{T}(\pi) \braket{\Phi | \hat E^{IJ}_{AB} \hat E^r_s \hat E^{ab}_{ij} | \Phi}_{\pi}\label{eq:DeccPrefactorEvalExample}
\end{align}
In this, and the upper and lower index lists are $U=[I,J,r,a,b]$, $L=[A,B,s,i,j]$, and the notation ``$\braket{\Phi | \hat E^{IJ}_{AB} \hat E^r_s \hat E^{ab}_{ij} | \Phi}_{\pi}$'' denotes the concrete contribution on the right hand side of Eq.~\eqref{eq:ExampleMatrixElement2} for a given permutation $\pi$.
As before, $\mathcal{T}(\pi)$ is the topological factor explained after Eq.~\eqref{eq:CcPrefactor}.
In explaining the transformation, we shall first assume the upper indices in $U$ to have mutually distinct numerical values; comments on the general case will be given after this discussion.

\emph{Step 1.} We first expand the string of spin-free excitation operators into spin orbital form.
   Since the upper indices are distinct, we can furthermore re-order the destruction operators such that contracted pairs stand next to each other (\emph{cf}.~\eqref{eq:CcPrefactor}), as long as sign factors are accounted for.
   For example, for the concrete permutation $\pi_x=[4,5,2,1,3]$, which yields $\pi_x(L)=[i,j,B,A,s]$, we would get:
   \begin{align}
      &\braket{\Phi | \hat E^{IJ}_{AB} \hat E^r_s \hat E^{ab}_{ij} | \Phi}_{\pi_x} \notag
   \\ &=
      \sum_{\sigma_1,\ldots,\sigma_5}  
      \langle \Phi | \hat e^{\dagger}_{I\sigma_1}
      \hat e^{\dagger}_{J\sigma_2} \hat e _{B\sigma_2} \hat e_{A\sigma_1}
      \hat e^{\dagger}_{r\sigma_3} \hat e _{s\sigma_3} \hat e^{\dagger}_{a\sigma_4}
      \hat e^{\dagger}_{b\sigma_5} \hat e _{j\sigma_5} \hat e_{i\sigma_4} | \Phi  \rangle.  \label{a1}
   \\&= {k(\pi_x)} \sum_{\sigma_1,\ldots,\sigma_5}   \langle \Phi | \left( \hat e^{\dagger}_{I\sigma_1} \hat e_{i\sigma_4} \right)
   \left(   \hat e^{\dagger}_{J\sigma_2} \hat e_{j \sigma_5} \right)    \left(   \hat{  e}_{B\sigma_2} \hat e^{\dagger}_{r\sigma_3}\right) \notag
   \\& \qquad\left(  \hat   e_{A\sigma_1} \hat e^{\dagger}_{a\sigma_4} \right) \left( \hat e_{s\sigma_3} \hat e^{\dagger}_{b\sigma_5} \right) | \Phi  \rangle. \label{a2}
   \end{align}
   Here the sign factor
   \begin{align}
      k(\pi)=\sign(\pi)(-1)^{n_\textrm{particle-co}} \label{eq:PhaseFactorSignForm}
   \end{align}
   encodes the number of operator transpositions required to reach this form Eq.~\eqref{a2} for a given permutation $\pi$.
   The $(-1)$ prefactors thereby originate from the anti-commutativity of Fermionic spin-orbital creation and destruction operators (with distinct indices), yielding a $(-1)$ factor any time two operators are exchanged; as there is one for each transposition in $\pi$, and one for each ``particle contraction'' (that is, each contraction of which the destruction operator stands to the left of the creation operator in the original expression), we directly obtain Eq.~\eqref{eq:PhaseFactorSignForm}.
   
\emph{Step 2.} In this form, each of the aligned creation-destruction pairs either reproduces the reference determinant or annihilates it, depending on whether $U(i)=L(\pi(i))$ and $\sigma_i=\sigma_{\pi(i)}$ or not. For example,
   \begin{align}
         &\langle \Phi | \hat e^{\dagger}_{I\sigma_1} \hat e_{i\sigma_5} = \delta^I_i \delta^{\sigma_1}_{\sigma_5} \bra{\Phi} .
   \end{align}
   This can be used to incrementally evaluate the coupling coefficient Eq.~\eqref{a2}, until we reach $\braket{\Phi|\Phi}=1$. Consequently,
   Eq.~\eqref{eq:DeccPrefactorEvalExample} can be evaluated into
   \begin{align}
      \langle \Phi | \hat E^{IJ}_{AB} \hat E^r_s \hat E^{ab}_{ij} | \Phi  \rangle = \sum_{\pi \in S_5}\mathcal{T}(\pi) k(\pi) \sum_{\sigma_1,\ldots,\sigma_5} \prod_{i=1}^{5} \delta^{U_i}_{L_{\pi(i)}} \delta^{\sigma_i}_{\sigma_{\pi(i)}}  \label{eq:PairedSpinSummation2}.
   \end{align}

\emph{Step 3.} The spin summation on the right-hand side of \eqref{eq:PairedSpinSummation2} can be evaluated in closed form:
   \begin{align}
      \sum_{\sigma_1,\ldots,\sigma_N} \prod_{i=1}^{N} \delta^{\sigma_i}_{\sigma_{\pi(i)}} = 2^{n_\textrm{cycles-in-$\pi$}}.\label{eq:SpinSumClosedFormCycles}
   \end{align}
   This is seen by considering that the factors $\delta^{\sigma_i}_{\sigma_{\pi(i)}}$ effectively collapse \emph{all} spin summations within a cycle in $\pi$ into a \emph{single} spin summation, which is then evaluated to give a factor of 2.
   Note that this also includes (trivial) cycles of length 1.
   In this example, cycles [4,1] and [3,5,2] contribute to a factor of 4, as 
   \begin{align}
& \left(  \sum_{\sigma_1 \sigma_4} \delta_{\sigma 1}^{\sigma_4} \delta_{\sigma_4}^{\sigma_1} \right)
\left(   \sum_{\sigma_2 \sigma_3 \sigma_5} \delta_{\sigma_2}^{\sigma_5} \delta_{\sigma_5}^{\sigma_3} \delta_{\sigma_3}^{\sigma_2} \right) \delta_i^I \delta_j^J \delta_B^r \delta_A^a \delta_s^b \nonumber \\
&= 4  \delta_i^I \delta_j^J \delta_B^r \delta_A^a \delta_s^b.
   \end{align}

\emph{Step 4.} The phase factor $k(\pi)=\sign(\pi)(-1)^{n_\textrm{particle-co}}$ from Eq.~\eqref{eq:PhaseFactorSignForm} can be shown to be equal to
\begin{align}
  k(\pi) = (-1)^{n_{\textrm{hole-co}}} (-1)^{n_{\textrm{cycles-in-$\pi$}}}.\label{eq:HolesParticlesCyclesResult}
\end{align}
This is seen as follows: 
First, let $n_\textrm{even-cycles-in-$\pi$}$ and $n_\textrm{odd-cycles-in-$\pi$}$ denote the number $\pi$'s cycles of even length and of odd length, respectively.
It is an elementary result from permutation group theory that for any permutation $\pi$, we have\cite{wilson2009finite}
\begin{align}
   \sign(\pi) = (-1)^{n_\textrm{even-cycles-in-$\pi$}}.
\end{align}
Furthermore, every point of $\pi$ is a member of exactly one cycle.
The total number of points in even-length cycles is therefore always even, and the total number of points in odd-length cycles is even if and only if there is an even number of odd-length cycles.
Therefore, for any permutation $\pi\in S_N$, we get
\begin{align}
   (-1)^{n_\textrm{odd-cycles-in-$\pi$}} = (-1)^N.
\end{align}
Combined, this yields
\begin{align}
   (-1)^{n_{\textrm{cycles-in-$\pi$}}} &= (-1)^{n_{\textrm{even-cycles-in-$\pi$}}}\cdot(-1)^{n_{\textrm{odd-cycles-in-$\pi$}}}
\\ &= \sign(\pi)\cdot(-1)^N.\label{eq:HolesParticlesCycles1}
\end{align}
Additionally, every contraction is either a hole contraction or a particle contraction, so
\begin{align}
   N = n_{\textrm{hole-co}} + n_{\textrm{particle-co}}.
\end{align}
Inserting this into Eq.~\eqref{eq:HolesParticlesCycles1} then yields
\begin{align}
   \sign(\pi)(-1)^{n_\textrm{particle-co}} = (-1)^{n_{\textrm{cycles-in-$\pi$}}} (-1)^{n_\textrm{hole-co}},
\end{align}
and thereby Eq.~\eqref{eq:HolesParticlesCyclesResult}.
Combining Eq.~\eqref{eq:HolesParticlesCyclesResult}, the prefactor from Fermionic operator alignment, with Eq.~\eqref{eq:SpinSumClosedFormCycles}, the prefactor from the spin summation, then directly yields the total prefactor $C(\pi)$ of Eq.~\eqref{eq:CcPrefactor}.

In summary, for the case of distinct numerical values of indices in $U$ (and therefore $L$), the form and factors of $C(\pi)$ in Eq.~\eqref{eq:CcPrefactor} can shown to be correct.
While only shown for one specific example term, the presented arguments can be readily generalized into a full proof of Eq.~\eqref{eq:CcPrefactor} for the general contraction case, as long as indices in $U$ remain distinct.
Unfortunately, this is not the case if non-distinct numerical values of indices in $U$ are present, as in this case we cannot proceed from the generalization of Eq. (\ref{a1}) to (\ref{a2}).
However, a variant of the argument used in traditional diagrammatic techniques (e.g., Ref.~\onlinecite{bartlett2009many} p.74ff), where exactly the same problem appears, can be used to establish that the \emph{form} of the resulting equations cannot actually depend on the concrete numerical \emph{values} of the indices in $U$.
For this reason, the presently outlined proof is sufficient to determine the values of the coupling coefficients.

\bibliography{topic,ref}

\end{document}